%% file: WISE_4FGL.tex
\documentclass[twocolumn]{aastex631}

\usepackage{hyperref}
\usepackage{natbib}
\usepackage{soul,comment}
\usepackage{color}
\usepackage{wrapfig}
\usepackage{numprint}
\usepackage{graphicx}

\shorttitle{BL Lacs and FSRQs in 4FGL}

\graphicspath{{./}{figures/}}

\begin{document}

\title{Using Neural Networks to Differentiate Newly Discovered BL Lacs and FSRQs among the 4FGL Unassociated Sources Employing Gamma-ray, X-ray, UV/Optical and IR Data}

\author[0000-0002-0878-1193]{Amanpreet Kaur}
\affiliation{Department of Astronomy and Astrophysics \\
 Pennsylvania State University
University Park, PA 16802, USA}

\author[0000-0003-2633-2196]{Stephen Kerby}
\affiliation{Department of Astronomy and Astrophysics \\
 Pennsylvania State University
University Park, PA 16802, USA}

\author[0000-0002-5068-7344]{Abraham D. Falcone}
\affiliation{Department of Astronomy and Astrophysics \\
 Pennsylvania State University
University Park, PA 16802, USA}

\begin{abstract}
Among the $\sim$ 2157 unassociated sources in the third data release (DR3) of the fourth $Fermi$  catalog, $\sim$ 1200 were observed with the Neil Gehrels $Swift$ Observatory pointed instruments. These observations yielded 238 high S/N X-ray sources within the 95\% $Fermi$  uncertainty regions. Recently, \citet{Kerby2021} employed neural networks to find blazar candidates among these 238 X-ray counterparts to the 4FGL unassociated sources and found 112 likely blazar counterpart sources. A complete sample of blazars, along with their sub classification, is a necessary step to help understand the puzzle of the blazar sequence and for the overall completeness of the gamma-ray emitting blazar class in the $Fermi$ catalog. We employed a multi-perceptron neural network classifier to identify FSRQs and BL Lacs among these 112 blazar candidates using the gamma-ray, X-ray, UV/optical and IR properties. This classifier provided probability estimates for each source to be associated with one or the other category, such that P$_{fsrq}$ represents the probability for a source to be associated with the FSRQ subclass. Using this approach, 4 FSRQs and 50 BL Lacs are classified as such with $>99\%$ confidence, while the remaining 58 blazars could not be unambiguously classified as either BL Lac or FSRQ. 

\end{abstract}

\keywords{}

\section{Introduction} \label{sec:intro}
The third release (DR3) of the fourth catalog \citep[4FGL, ][]{2022-4FGL}) of $Fermi$  Large Area Telescope (LAT) catalog with 6658 sources contains approximately 2157 unidentified/unassociated sources (i.e. approximately one-third are unassociated). The lack of identification for so many sources represents a major gap in our understanding of the high energy universe. Therefore, exploring the nature of these sources remains fundamental to understanding the gamma-ray universe. Most of the identified objects in the catalog are blazars ($\sim$75\%), and the others are pulsars, X-ray binaries, supernova remnants, novae, GRBs etc. The brighter of these blazars offer the opportunity to study the more extreme gamma-ray emitters on an individual basis, while the lower-significance and dimmer new blazars offer the opportunity to explore the population in a more complete manner by extending the discovery space to lower fluxes. It is likely that the unassociated source population contains a significant number of blazars, which are missing associations for various reasons such as being on the fainter side or extreme nature etc. In general, blazars are defined as a powerful subclass of the Active Galactic Nuclei which have their jets pointing along our line of sight \citep{Blandford1978}. The overall spectral energy distribution of these sources display a double-bump structure, such that the synchrotron emission process is responsible for the lower energy bump (Radio to X-ray) and the higher  one (X-ray rays to gamma-rays) is attributed to either synchrotron self Compton mechanism and/or external inverse Compton processes, and/or hadronic processes such as proton synchrotron \citep[e.g., see][]{Bottcher2013,2019VandenBerg}. 
Blazars are further classified into two subclasses based on their optical spectra; Flat Spectrum Radio Quasars (FSRQs) and BL Lacertae Objects (BL Lacs). The former exhibit broad emission lines, whereas the BL Lacs show no lines or narrow lines (equivalent width $<$ 5 \AA) \citep{1991Stickel}. 
Understanding the origin of the differences between the two subclasses of blazars has been one of the open questions in the field. The idea of the blazar sequence, first described by \citet{Fossati1998} and re-visited by \citet{Ghisellini2008, Ghisellini2017}, revealed that the most luminous blazars possess lower synchrotron frequencies, which often is the case in FSRQs, and vice versa for BL Lacs. Based on this principle, \citet{2010ApJ...715..429A} divided blazars
into three categories based on their  synchrotron peak frequencies ($\nu^{pk}_{syn}<$) as follows: low-synchrotron-peaked blazars; LSP: $\nu^{pk}_{syn}<$ 14,  ISP: $14 > \nu^{pk}_{syn}>$ 15 and HSP: $\nu^{pk}_{syn}>$ 15 such that most of the FSRQs belong to the LSP category and BL Lacs contribute to the ISP and HSP. However, \citet{2012Giommi} argue that the blazar sequence can be an effect of selection bias and also this sequence has been challenged by finding highly-luminous, high-synchrotron-peaked blazars, e.g. \citet{Padovani2012}. Moreover, \citet{Ghisellini1998} suggested a unified scheme for blazars in which FSRQs eventually evolve into BL Lacs once their accretion disk is exhausted. The observational evidence of finding FSRQs at typically higher redshifts than that of BL Lacs supports this scenario. In the past few years, several ($\sim$30) BL Lacs have been found at high redshifts e.g., \citep{Rau2012, Kaur2017, Kaur2018, Rajagopal2020, 2020Paliya,2022ApJ...929L...7K}. The blazar parameter space and the theories regarding their evolution need to be further explored by obtaining a more complete sample of both types of blazars. As an example, also see \citet{Paliya-central2021} to see bimodality in blazars based on the disk luminosity. Therefore, the rather large unassociated 4FGL source list needs to be probed to find additional blazars, which then need to be classified as FSRQs or BL Lacs using spectroscopic observations. Ideally one would also obtain and analyze broadband multiwavelength data for these gamma-ray sources to identify and evaluate possible counterparts, which would be take multiple years for completion. \\
The Neil Gehrels Swift observatory ($Swift$) has completed X-ray and UV/optical observations for $\sim$ 1000 unassociated sources through a fill-in program \citep{Falcone2020} based on the following criteria: (i) the 95\% $Fermi$ uncertainty region associated with the source is less than the Swift-XRT field of view, which is 23 arcmin across, (ii) the source was not on the galactic ridge where diffuse background could lead to false associations. With a typical exposure time of $\sim$4\,ksec, these observations resulted in a total of 238 X-ray counterparts found (with S/N $\ge$ 4) within the 95\% uncertainty regions of 208 $Fermi$ 4FGL-DR3 sources. The photometric data including X-ray images, detected sources, count rates, SNR etc. for all of these unassociated sources are available on the public web page \href{http://www.swift.psu.edu/unassociated/}{http://www.swift.psu.edu/unassociated/}.

In the past few years, various authors explored using machine learning to classify blazars such as \citet{Doert2013, Salvetti2017, Kang2019, Kaur2019a, Kaur20193BzPsr, Kovacevic2020, Chiaro2021}.
Most recently, \citep{Kerby2021} (K21 from now on) has found blazar as well as pulsar candidates among these 238 X-ray sources employing the gamma-ray, X-ray and UV/Optical properties using a method of neural networks. The objective of this letter is to classify these blazar candidates further into their subclasses; FSRQs and BL Lacs using a neural network classifier employing the data from gamma-ray, X-ray, UV/Opt, and IR regimes. This letter is organized as follows:
Section~\ref{sec:obs} describes the multi-wavelength sample selection criteria for this work, Section~\ref{sec:analysis} explains the analysis procedure for classification, and Section~\ref{sec:results} describes the results. The concluding remarks, as well as discussion in regard to the results obtained, are provided in Section~\ref{sec:conclusions}.

\section{Observations and Data Selection} \label{sec:obs}
\subsection{Sample Selection - Blazar Candidates }
The source sample for this work is derived from the work of K21, therefore the basic analysis steps are only briefly described in this section. The thorough details of each step for analysis are already explained in K21. 

\begin{figure*}
    \centering
    \includegraphics[width=2\columnwidth]{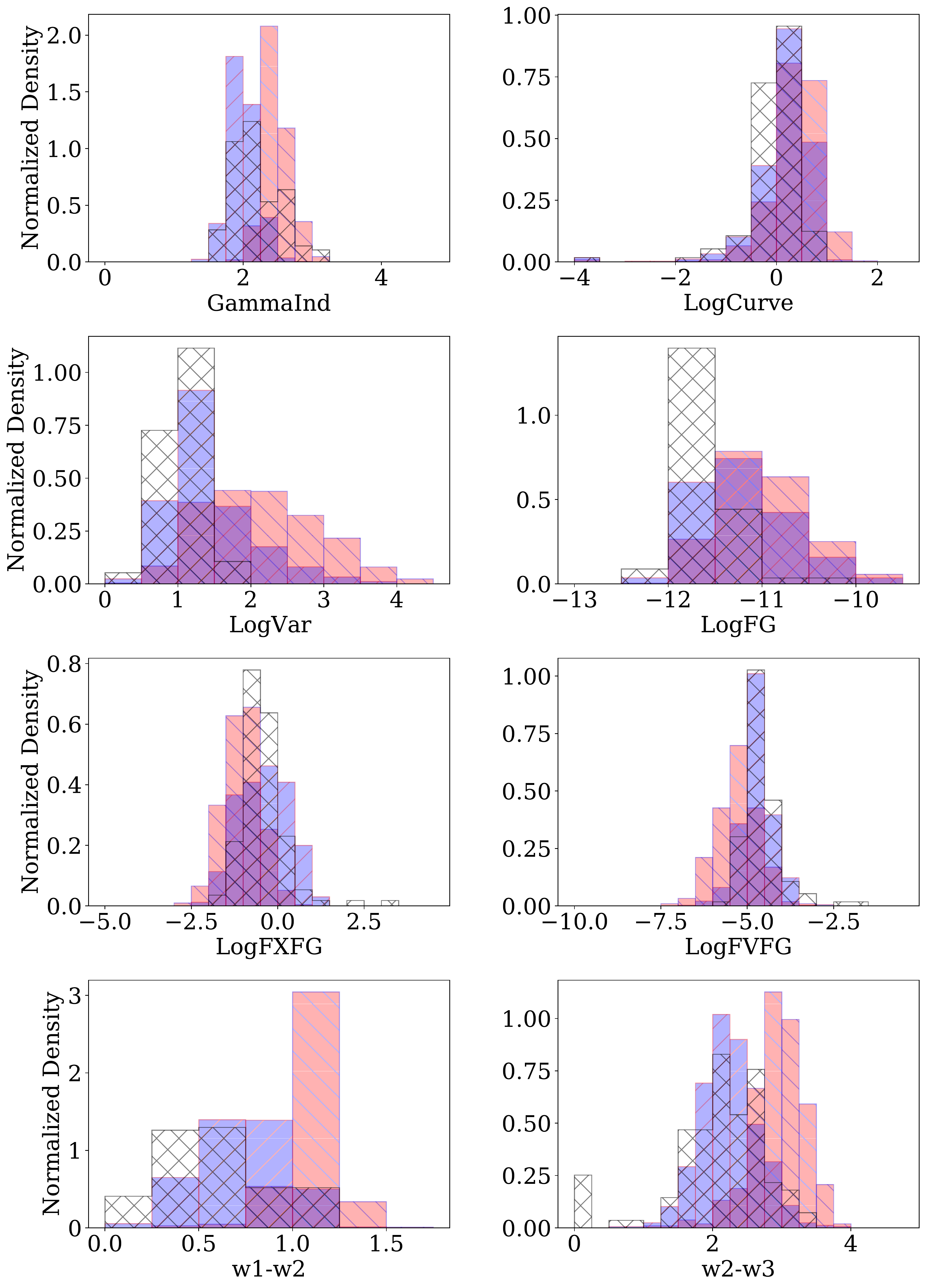}
    \caption{Normalized histograms for the known BL Lacs (blue), known FSRQs (red), and unassocicated (black line) samples.}
    \label{fig:AllHistos}
\end{figure*}

From the list of $\sim$ 2157 4FGL unassociated sources, the Neil Gehrels Swift Observatory ($Swift$) had completed observations for $\sim$ 500 sources by February 1, 2021  with an average exposure time of 4\,ksec per source to find potential X-ray counterparts. A total of 238 X-ray sources with SNR$\ge$4 were found among the 208 4FGL unassociated fields within their 95\% uncertainty regions. Out of these, more than one X-ray source was found within the 95\% regions of the 14 4FGL fields, whereas the rest have exactly one X-ray source within this region. The basic photometric results such as the \texttt{X-ray Right Ascension, X-ray Declination, Count Rate and SNR} are made available to the community on the webpage\footnote{https://www.swift.psu.edu/unassociated}. K21 collected these data for 238 X-ray sources and conducted X-ray spectral analysis fitting \verb|powerlaw| models using \verb|XSpec v12.10|\citep{Arnaud1996} to obtain spectral parameters such as spectral indices and X-ray fluxes in the 0.3-10\,keV band. A thorough description of the analysis procedure is provided in Section 3.1 in K21. 
Whenever $Swift$ observes  using the X-ray telescope, observations are also conducted with at least one out of the six UV-optical filters (v, b, u, uvw1, uvm2, uvw2; ranging from 5468\AA to 1928\AA) mounted on the UltraViolet Optical Telescope (UVOT)\citep{Roming2005}. For any UV/optical source(s) found within the X-ray positional uncertainty regions, a photometric analysis was conducted using \verb|uvotsource| task in the  $Swift$ data analysis pipeline which provides the magnitudes in the AB system. Whenever no source was found, an upper limit was calculated. In a few cases, there were indistinguishable multiple sources within the X-ray positional uncertainty regions, these sources were removed from our final list of uassociated sources. All the details of this analysis are provided in Section 2.2 in K21. Finally, these magnitudes were then converted into equivalent of the Johnson V magnitudes for uniformity across the source list, the details of this conversion as well as all the assumptions are clearly laid out in section 3.2 in K21. In the end, the final sample comprised 205 sources corresponding to 188 4FGL unassociated fields for which X-ray as well as UV-opt data were available. Out of these, 174 had single X-ray and UV sources in their respective 4FGL uncertainty regions and 14 had multiple X-ray/UV sources. \\
For these 205 sources, K21 conducted a neural network analysis to find that 132 were likely blazars with probabilities, P$_{bzr}$, $\ge$99\%, where P$_{bzr}$ is the probability for a source to be associated with a blazar class as provided by the underlying neural network method. The two subclasses of the gamma-ray emitting blazars display distinct colors when compared in a 2-dimensional color-color plot using the publicly available data, AllWISE\citep{Cutri2013} obtained with the Wide-field Infrared Survey Explorer (WISE) \citep{Wright2010}. These authors plotted color indices, \verb|w1-w2| vs \verb|w2-w3|, (\verb|w1, w2, w3| correspond to 3.4, 4.6, 12\,$\mu$m, respectively)  for various known gamma-ray emitting BL Lacs and FSRQs and found that these two sublcasses occupy different parameter space \citep{Massaro2011,Massaro2012,DAbrusco2014}. Since the objective of the work presented here is to classify the blazar candidates into BL Lacs and FSRQs, we employ the infrared properties along with existing gamma-ray, X-ray and UV/opt data for these sources. In order to do so, we conducted WISE positional matches for the $Swift$-UVOT positions within 5'' (maximum XRT positional uncertainty for a given source in our sample) for the 132 blazar candidates and found 112 matches. We note that it is possible that a few of these WISE counterparts could be spatially coincident with the XRT and UVOT position by a random chance, but the probability of this is low. In order to confirm this, we used the $Swift$-XRT blazar catalog with 2831 blazars using 15 years of data \citep{Giommi2019}. When searched for WISE counterparts within 5'' positional uncertainities, it was found that only 22 secondary sources were found in addition to the actual WISE counterparts for these blazars. This suggests a  $\sim$ 0.7\% probability for a WISE source which can be found via chance coincidence at a random position in the sky within a 5'' uncertainty region. We, therefore, finalize our source sample size of blazar candidates among the 4FGL unassociated source list to 112 such that the required gamma-ray, X-ray, UV-opt and IR properties are known for these sources. The relevant eight properties for this classification work are as itemized below. The parameters for each training or research source include:
\begin{itemize}
    \item Gamma-ray photon index, $\Gamma_\gamma$ (\verb|PL_Index| in the 4FGL catalog)
    \item The logarithm of gamma-ray flux, $\log{F_\gamma}$, in $\rm{erg/s/cm^2}$ (\verb|Energy_Flux| in the 4FGL catalog)
    \item The logarithm of X-ray to gamma-ray flux ratio, $\log{F_X/F_\gamma}$
    \item The logarithm of V-band to gamma-ray flux ratio, $\log{F_V/F_\gamma}$.
    \item The significance of the curvature in the gamma-ray spectrum, henceforth simply \textit{curvature} (\verb|PLEC_SigCurv| in the 4FGL catalog)
    \item The year-over-year gamma-ray variability index (\verb|Variability_Index| in the 4FGL catalog)
    \item WISE color index 1, \verb|w1-w2|
    and
    \item WISE color index 2, \verb|w2-w3|
\end{itemize}
Normalized histograms of all parameters for both the known FSRQs and BL Lacs and the research sample are shown in Figure \ref{fig:AllHistos}. 

\subsection{Training Data: Known BL Lacs and FSRQs}
In order to classify our sample of blazar candidates into the two subclasses; BL Lacs and FSRQs, we would require a large sample of these two categories for which the exact same gamma-ray to IR properties can be established. The neural network described later would use the information from this known sample of two classes to train itself to be able to distinguish between them. For this step, we selected a sample of 427 FSRQs and 671 BL Lacs for which the relevant multiwavelength data, i.e., the above mentioned eight properties were available. For these blazars, all the gamma-ray properties were obtained from the 4FGL catalog, X-ray fluxes were provided in the BZCAT \citep{Massaro2015}, which were obtained from the ROSAT X-ray source catalog\citep{Voges1999}. These X-ray fluxes were provided in the 0.1-2.5 keV band, but these were converted to 0.3-10 keV for consistency using the Portable, Interactive Multi-Mission Simulator \verb|PIMMS|\citep{Mukai1993}. The photon indices were assumed to be 2.0 and the hydrogen column densities were obtained using the LAB survey\citep{Kalberla2005} for the conversion process. The WISE data were also obtained from the BZCAT catalog. The optical fluxes were obtained from various resources such as SIMBAD Astronomical Database\footnote{http://simbad.u-strasbg.fr/simbad/}, the Sloan Digital Sky Survey\footnote{https://www.sdss.org} as well as Gaia archive\footnote{https://gea.esac.esa.int/archive/}. All the magnitudes obtained were converted to V band using the same principles as outlined in section 3.2 in K21, which were then converted to the respective optical flux values. Once the training data as well as the unknown data (blazar candidates without their subclassification) are finalized, the next step was to develop a neural network classifier which is explained below. The eight properties for all the sources in the source sample (112) as well as the training sample (1098) are provided in Table~\ref{tab:sample} and Table~\ref{tab:bzrs}, respectively. It should be noted that variability in blazars is witnessed in both the subclasses but it is more pronounced in the FSRQs as compared to BL Lacs possibly due to the brightness factor (FSRQs are brighter and easily detectable). This effect is also seen in our training sample which therefore is an efficient parameter for the classification. Even though the Fermi data used for this work are taken and averaged over multiple years as compared to the X-ray, UV and WISE data which were obtained within a few hours, the signal-to-noise ratio or data quality for all these lower energy regimes taken over a few hours is comparable to Fermi data taken over years. Therefore, using multiwavelength data for this classification is justified.

\section{Analysis-Neutral Network Classification
}\label{sec:analysis}
\begin{figure}[t]
    \centering
    \includegraphics[width=\columnwidth]{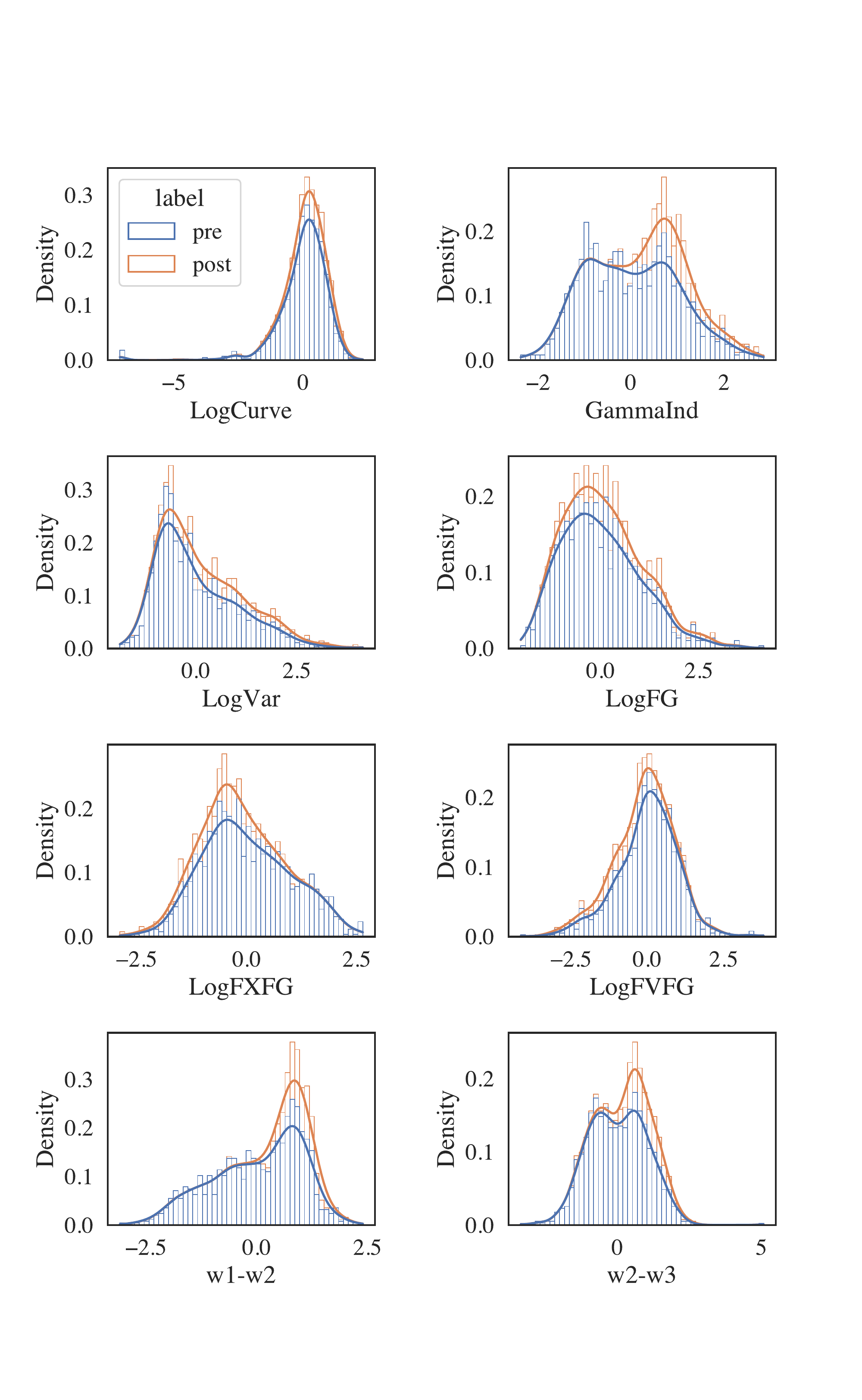}
    \caption{A comparison of the distributions of BL Lacs and FSRQs for each parameter before and after applying SMOTE. It is clearly seem that the change in distributions are
    minimal. The label \texttt{pre} corresponds to data before using SMOTE to artificially add more data and \texttt{post} after the application to balance both the classes.
    }
    \label{fig:smote}
\end{figure}

To classify the research sample of new likely blazars into likely FSRQ and likely BL Lacs, we applied a multi-layer perceptron (MLP) neural network classifier (NNC) approach. Because the number of known FSRQs and BL Lacs in our catalogs were slightly unbalanced (427 known FSRQs and 671 known BL Lacs), we first applied Synthetic Minority Over-sampling Technique (SMOTE) \citep{Chawla2002} to generate additional FSRQs to create a balanced dataset made up of real FSRQs and BL Lacs plus additional FSRQs created with a k-nearest neighbors approach to match the distribution of real FSRQs. It is essential to have a larger training sample for better accuracies for the classification, therefore a technique like SMOTE is used to increase the overall sample size instead of decreasing it. The resulting post-SMOTE distributions for all the parameters are similar to the pre-SMOTE distributions, see Figure~\ref{fig:smote}.\\

To build the NNC for this research, we split the catalog of known FSRQs and BL Lacs (plus SMOTE-generated FSRQs) into a training subsample and a validation subsample in an 80/20 split with proportional amounts of FSRQs and BL Lacs in each subsample. Building the NNC method, we constructed an approach where the unassociated sources are assigned a FSRQ probability $P_{fsrq}$. The NNC has eight input nodes (one for each parameter of the training dataset), one hidden layer with three neurons, and one output node returning the predicted FSRQ probability. In this research we use the \verb|MLPClassifier| function of the \verb|scikit-learn| Python package for NNC training and validation, training the NNC with the `adam' optimization approach \citep{Kingma2014}.

To determine when to stop iterating while training the NNC, we used the \textit{Log-Loss} parameter, an error measure in binary classification approaches that is similar to a likelihood estimator. For a sample with true classifications $y_i = (0,1)$ and predicted classification $p_i \in [0,1]$, the Log-Loss is given by

$$ L_{log} = - \sum_i \left( y_i \log(p_i) + (1-y_i) \log(1-p_i) \right)$$

At each iteration step in training the NNC, we calculate $L_{log}$ of both the training and validation subsamples.  As training continues, the training subsample $L_{log}$ should continuously decrease as the NNC approaches a more exacting fit. However, at some point the validation $L_{log}$ begins to increase as the NNC starts to overfit the training sample and lose predictive capabilities on unknown data points.  At this point, training is stopped. 
\begin{figure}
    \centering
    \includegraphics[width=\columnwidth]{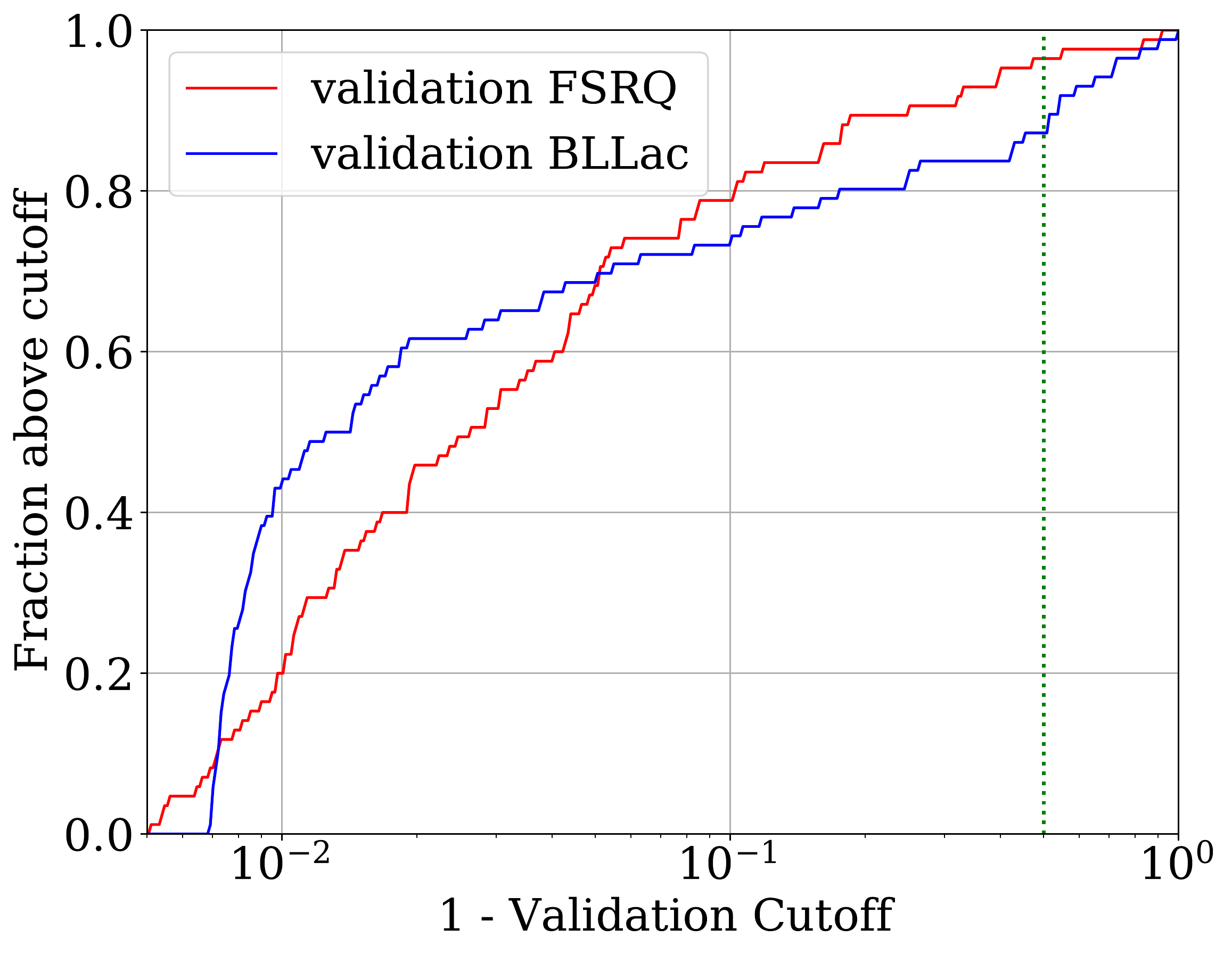}
    \caption{The fraction of the validation FSRQ (red) and BL Lac (blue) correctly classified as such with various cutoff limits considered a success. This plot also serve as a CDF-like representation of the validation $P_{fsrq}$ for each subclass. The green dotted line is a 50\% cutoff. We interpret the two slopes in this plot as the ability of the neural network to converge to a probability value for a given cut off in both the source classes. This plot shows that more BL Lacs converge to higher probabilities as compared to FSRQs, which we also see in our results.}
    \label{fig:NNCcdf}
\end{figure}

To evaluate the accuracy of the NNC, we applied the trained classifier to the validation dataset of known FSRQ and BL Lac objects.  Instead of establishing a single cutoff for FSRQ/BL Lac classification, we create a cumulative distribution function to investigate the fraction of known FSRQs and BL Lacs classified with various confidences.  For example, a perfect classification approach would assign all validation FSRQs $P_{fsrq} = 1$ and all validation BL Lac $P_{fsrq} = 0$. The CDF-like plot produced by comparing these ideal scores to the validation $P_{fsrq}$ assigned by the NNC is shown in Figure \ref{fig:NNCcdf}. The validation FSRQ and BL Lac are considered correctly classified if $P_{fsrq} > 0.5$ or $< 0.5$ respectively, then our NNC achieves a validation score of over 90\% for both subclasses. If we require 90\% confidence in our classifications, the NNC achieves an 80\% accuracy for validation BL Lac and almost 90\% accuracy for FSRQ.

\section{Results}\label{sec:results}
\subsection{Classification Results}
While the sources with $P_{fsrq} > 0.5$ are more likely to be FSRQs and $P_{fsrq} < 0.5$ are more likely to be BL Lacs, we adopt a more conservative approach when classifying; only the sources for which the $P_{fsrq} > 0.99$ are considered to be \texttt{highly likely FSRQs} and only sources with $P_{fsrq} < 0.01$ are classified as \texttt{highly likely BL Lacs}. All the other sources are termed as \texttt{ambiguous-type blazar candidates}. The validation score for both the subclasses is shown in Figure~\ref{fig:NNCcdf}. This analysis resulted in 50 sources as  highly likely BL Lacs and 4 as highly likely FSRQs. In other words, we found 50 sources for which $P_{fsrq} < 0.01$ and 4 for which $P_{fsrq} > 0.99$. The remaining 58 sources are labelled as being blazars whose type is ambiguous. These results are shown in Table~\ref{tab:results}.
In addition, we conducted literature search to look for any known blazars or other sources at the UVOT positions for all these sources. We found 19 additional blazar candidates, 10 BL Lacs and 9 FSRQs(QSOs), 2 Seyferts, one cataclysmic variable, one rationally variable star as well as a few radio galaxies. Out of the 4 NN identified FSRQs, 2 are independently confirmed by this literature study.  This literature search was conducted using the SIMBAD Astronomical Database \citep{Wenger2000} as well the latest data release (DR16) for the Sloan Digital Sky Survey database \citep{SDSS2020}. All these positional coincidences which yielded source classifications as well redshifts estimates are noted in Table~\ref{tab:results}. 
\begin{figure}[t]
    \centering
    \includegraphics[width=\columnwidth]{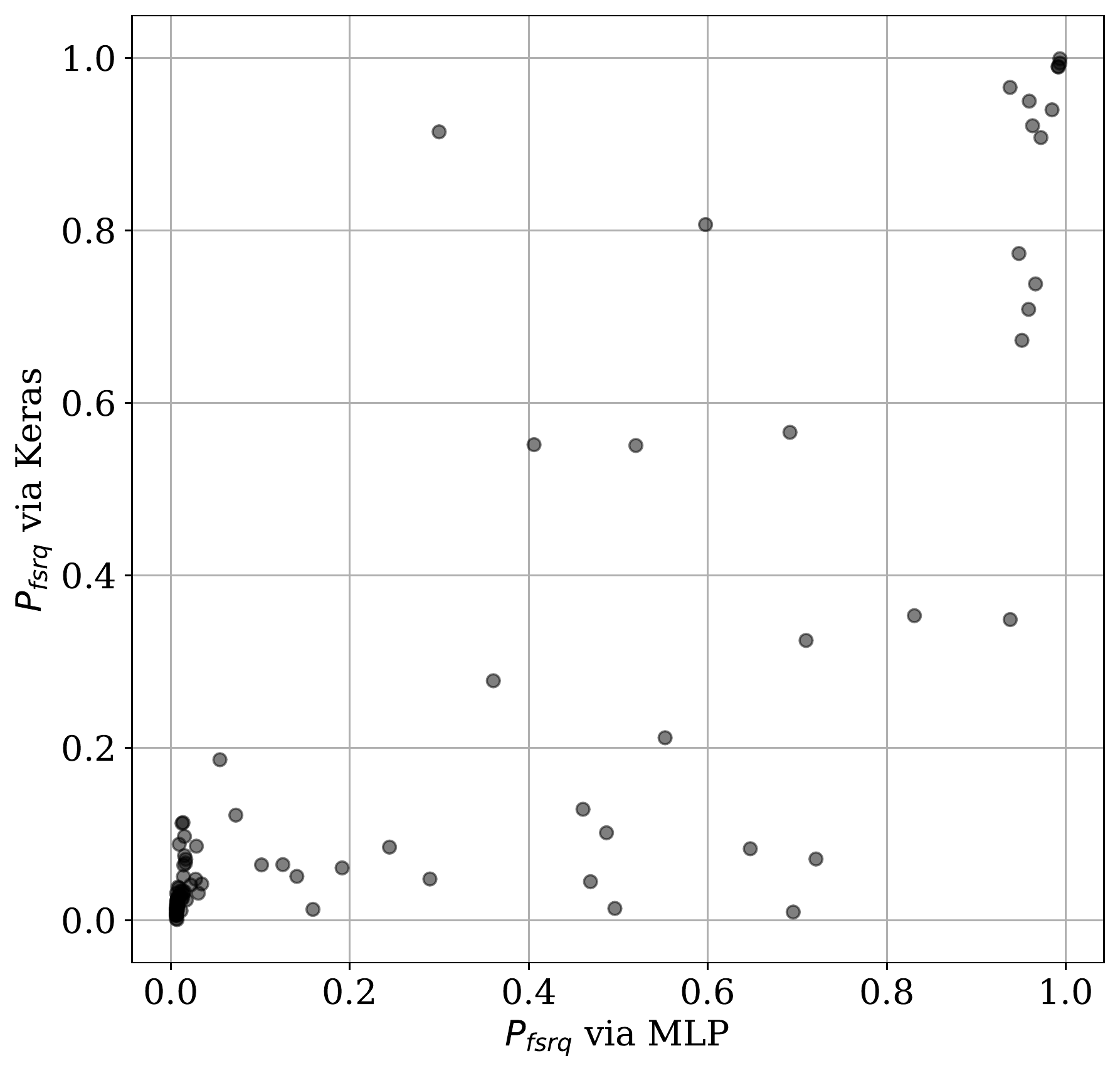}
    \caption{A comparison of the probabilities for a blazar to be an FSRQ, $P_{fsrq}$ is shown resulting from the two neural network approaches; MLP and Keras approach. It is clearly evident that the resulting highly likely FSRQs and BL Lacs are consistent with both the approaches, thereby proving the results of this work independent of a NNC approach.}
    \label{fig:comparethetwo}
\end{figure}
\subsection{Verification of the results}
To verify the stability of our NNC classification scheme, we returned to the validation step to ensure that our selection of random seed, NNC structure, and validation subsample did not unexpectedly bias our final classifications. To examine this possibility, we re-validated our neural network in three ways.

First, we compared our NNC classification of the unassociated sources with another NNC approach, which employs batch normalization as well as dropping out layers in order to make the training network more stable and train faster. The process of batch normalization is analogous to normalization for a dataset, where the data in a given layer is normalized with respect to its own mean and standard deviation. This helps in the overall optimization process since this normalized data would need fewer epochs to complete training the underlying model \citep{pmlr-v37-ioffe15} thereby making the overall training process efficient. Furthermore, adding dropout reduces the risk of overfitting, which is generally caused by the network learning spurious patterns in the training data. By introducing a random dropout for a given fraction of a layer's input units, the chance of having these spurious patterns involved in the overall training decreases significantly \citep{JMLR:v15:srivastava14a}. 
We applied the \verb|keras| Application Programming Interface (API) from the \verb|tensorflow ver 2.6.0| library in \verb|python 3.9.7| to construct as well train this model. This \texttt{keras NNC} consisted of eight neurons (one for each underlying parameter for providing distinction between the two subclasses) which were batch normalized, this was followed by one hidden layer with five neurons and the activation function \verb|relu|. After this initial training, a dropout layer with a standard/default rate of 30\% was introduced followed by another batch normalization process and finally the output was generated with one neuron and the activation function \verb|sigmoid|. \\
The training procedure employed the \verb|adam| optimizer as well as the $Log$-$Loss$ parameter just in the same way as in the case of the MLP classifier (see Section~\ref{sec:analysis}) to create the final model to be applied to the source sample. In other words, we used similar optimization as well training procedure but employing a completely different method to test the dependability of our results on these different approaches. If the MLP approach we used provides comparable $P_{fsrq}$ to the Keras approach, we gain confidence in our classifications. As Figure \ref{fig:comparethetwo} shows, the unassociated sources have very similar classifications in the two approaches, suggesting that our classification results are not overly sensitive to NNC structure. It should be noted that we emphasize the robustness of classifications in this context, which is based on only the high confidence probabilities for each class; i.e.,  the $P_{fsrq}$ is either $\ge$ 0.99 or $\le$ 0.01 for FSRQs or BL Lacs, respectively. The vast majority of sources that are graded as likely FSRQ or BL Lacs are so classified in both approaches. We also note that any two neural network methods would not yield exactly the same results since we are using probabilities to classify the underlying source class. However, it is important that the high confidence source classification should be consistent regardless of the method employed. In this case, both the methods are consistent with each other when we confidently classify a source to be a BL Lac or an FSRQ only when their respective probabilities are greater than or equal to 99\%. All the sources classified with $>$99\% BL Lac probabilities using Keras method are also classified as BL Lacs using MLP.\\
Our work does not classify a source as a BL Lac or an FSRQ with their respective probabilities below 99\% to minimize misclassification of these two subclasses.

\begin{figure}[t]
    \centering
   \includegraphics[width=\columnwidth]{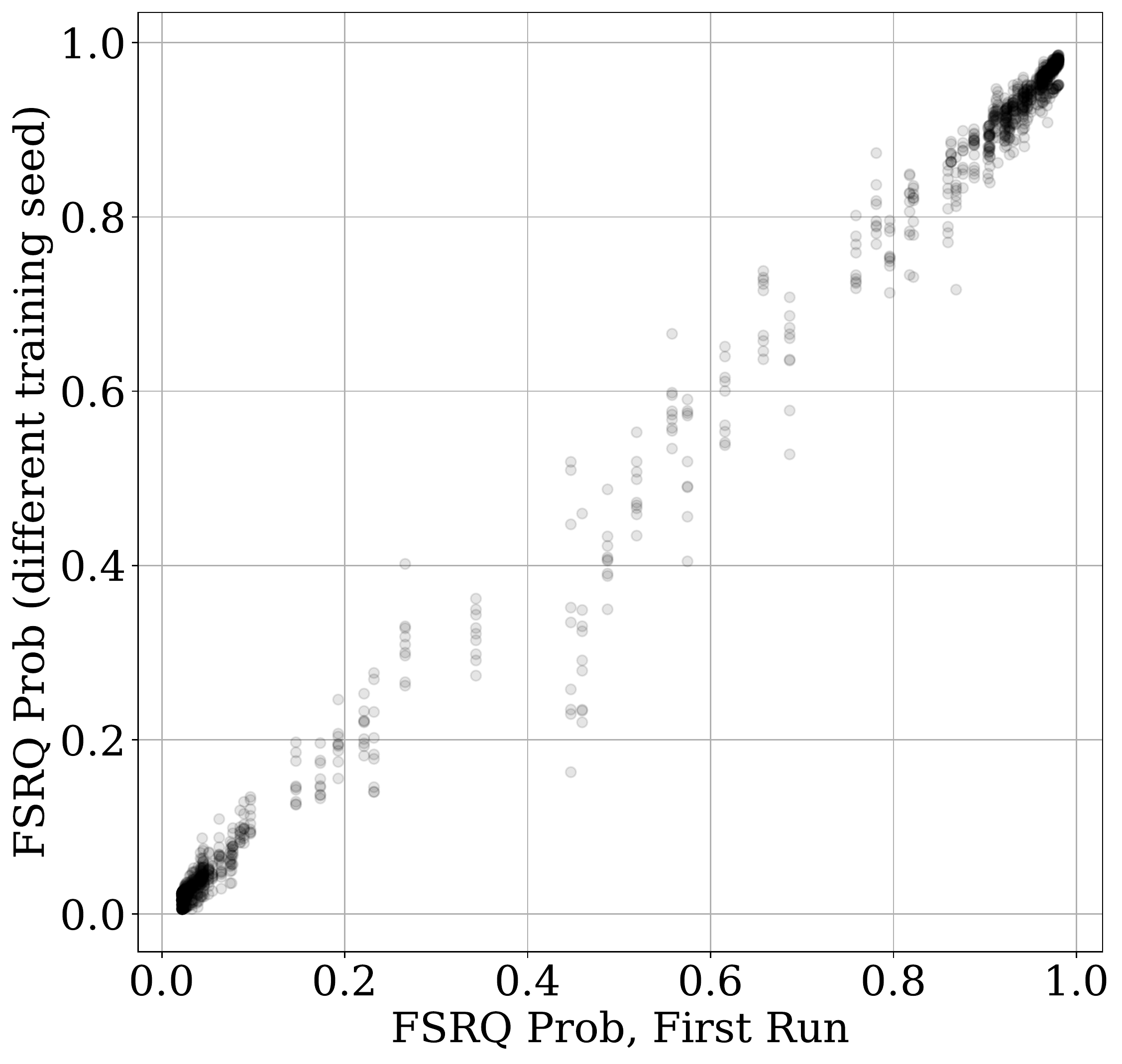}
    \caption{Keeping the same training and validation subsamples, the resulting $P_{fsrq}$ values with different random seeds for NNC training are shown here. It is clearly seen from the figure that the results are stable/consistent irrespective of the choice of the underlying random seed, in particular, for the highly likely FSRQs as well as BL Lacs.}
    \label{fig:SeedVary}
\end{figure}

Next, we varied the random seed of the training algorithm for the MLP classifier through a range of values, executing the training step with the same validation subsample. In this way, we can compare the validation $P_{fsrq}$ values on a single validation dataset while varying only the random seed of the NNC training.  Figure \ref{fig:SeedVary} compares the $P_{fsrq}$ score of the first of these iterations $i=0$ with nine others up to $i=10$. There are only small changes in the resultant $P_{fsrq}$ values from one iteration to the next, so we have confidence that our selection of random seed for NNC training did not bias our classification results.

\begin{figure}[b]
    \centering
    \includegraphics[width=\columnwidth]{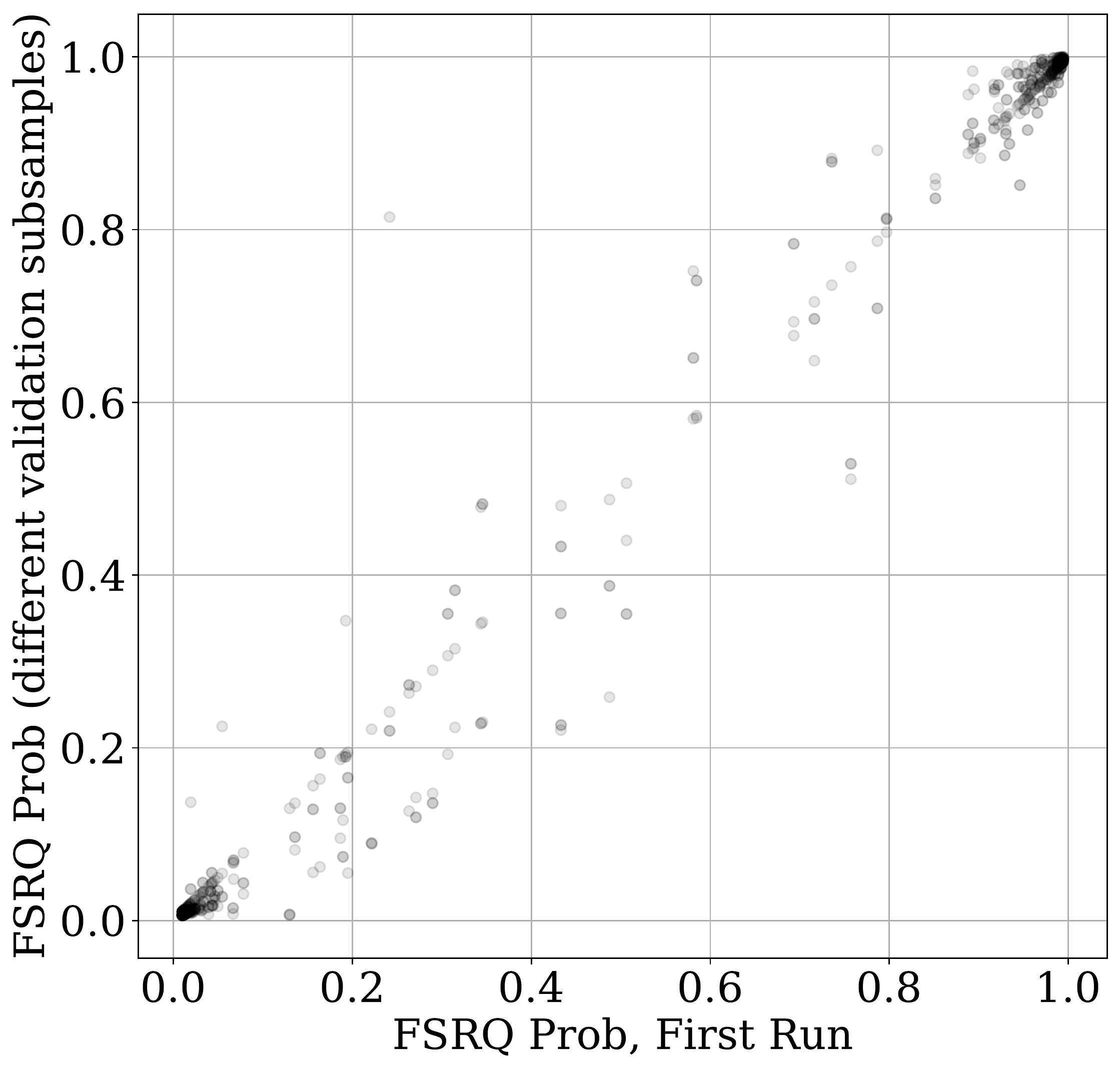}
    \caption{With a super-validation sample set aside, we compare $P_{fsrq}$ results for a super-validation sample, a proxy for our research sample, while varying which fold of the remaining known blazars serves as validation subsamples.}
    \label{fig:FoldVary}
\end{figure}

Finally, we examine whether our specific division of the collection of known FSRQs and BL Lacs into specific training and validation subsamples impacted final classification results. First, we split off 20\% of the known blazars into a `super-validation' subsample. With this list serving as a proxy for the research sample, we split the remaining known pulsars and blazars into quintiles, then train five NNCs. Each NNC uses one quintile as a validation subsample and the remaining as training subsample, akin to a stratified fold cross-validation approach. After training these five NNCs, we apply each to the super-validation sample, and compare $P_{fsrq}$ scores between iterations. Figure \ref{fig:FoldVary} shows that changing the validation subsample has little effect on the $P_{fsrq}$ results for a proxy research sample, and therefore we have little reason to suspect that our research results are sensitive to randomness in that way.
\section{Discussion and Conclusions}\label{sec:conclusions}
\begin{figure*}
\gridline{\fig{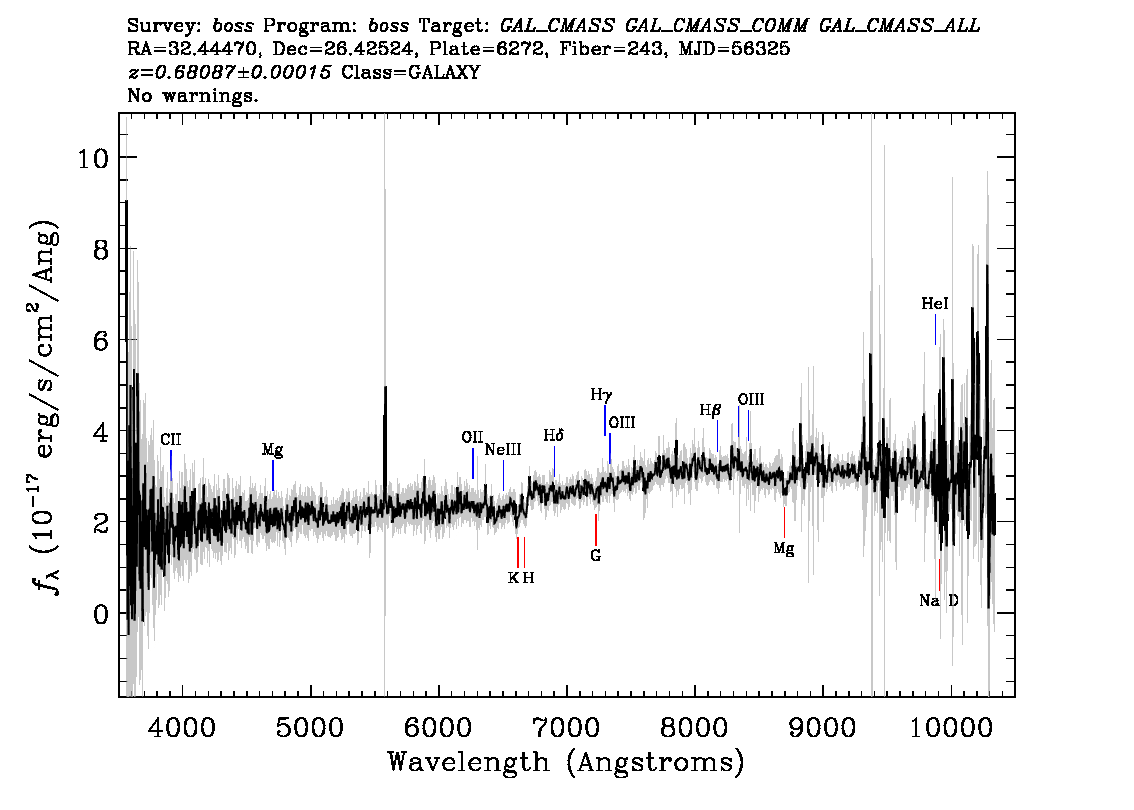}{0.5\textwidth}{(a)}
          \fig{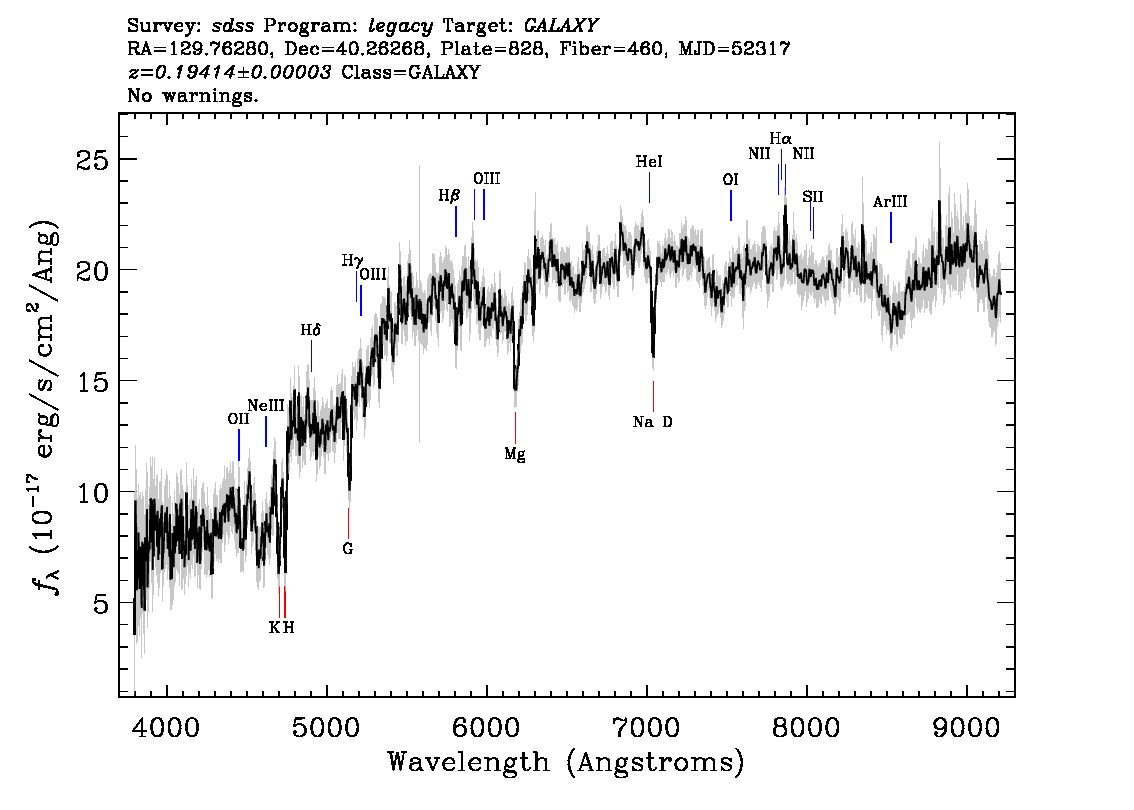}{0.5\textwidth}{(b)}
          }
\gridline{\fig{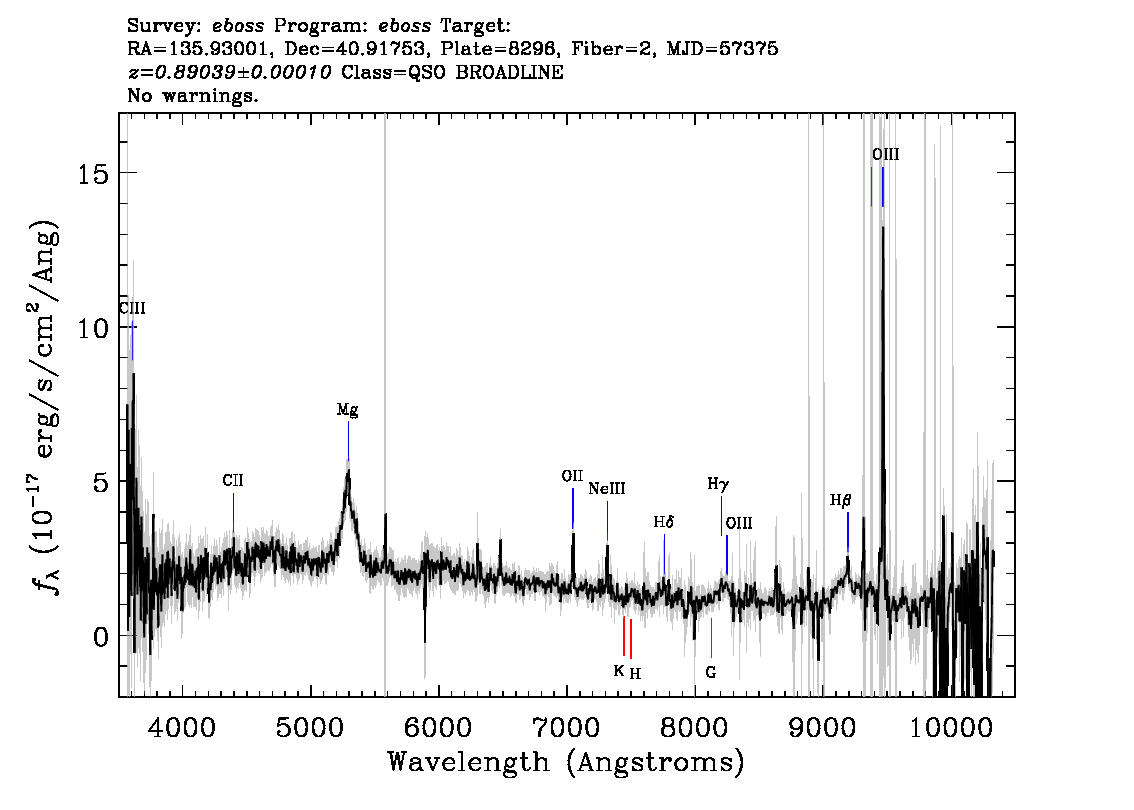}{0.5\textwidth}{(c)}
          \fig{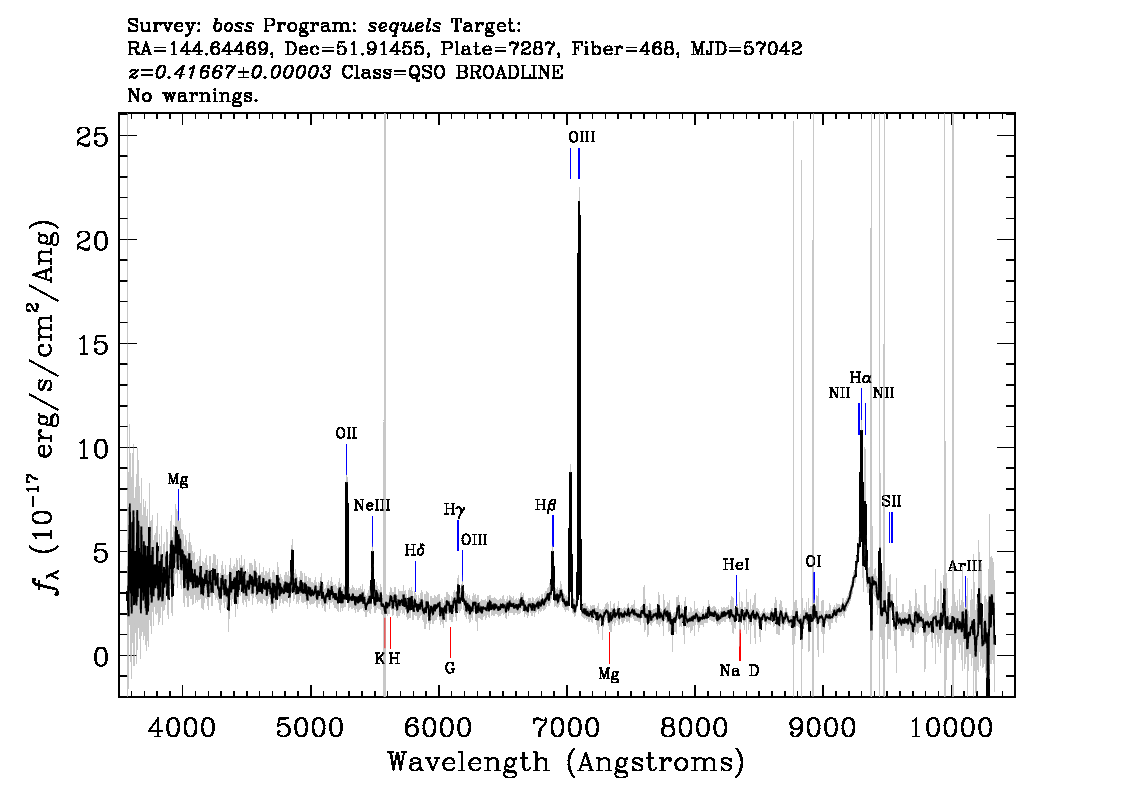}{0.5\textwidth}{(d)}
          }
\gridline{\fig{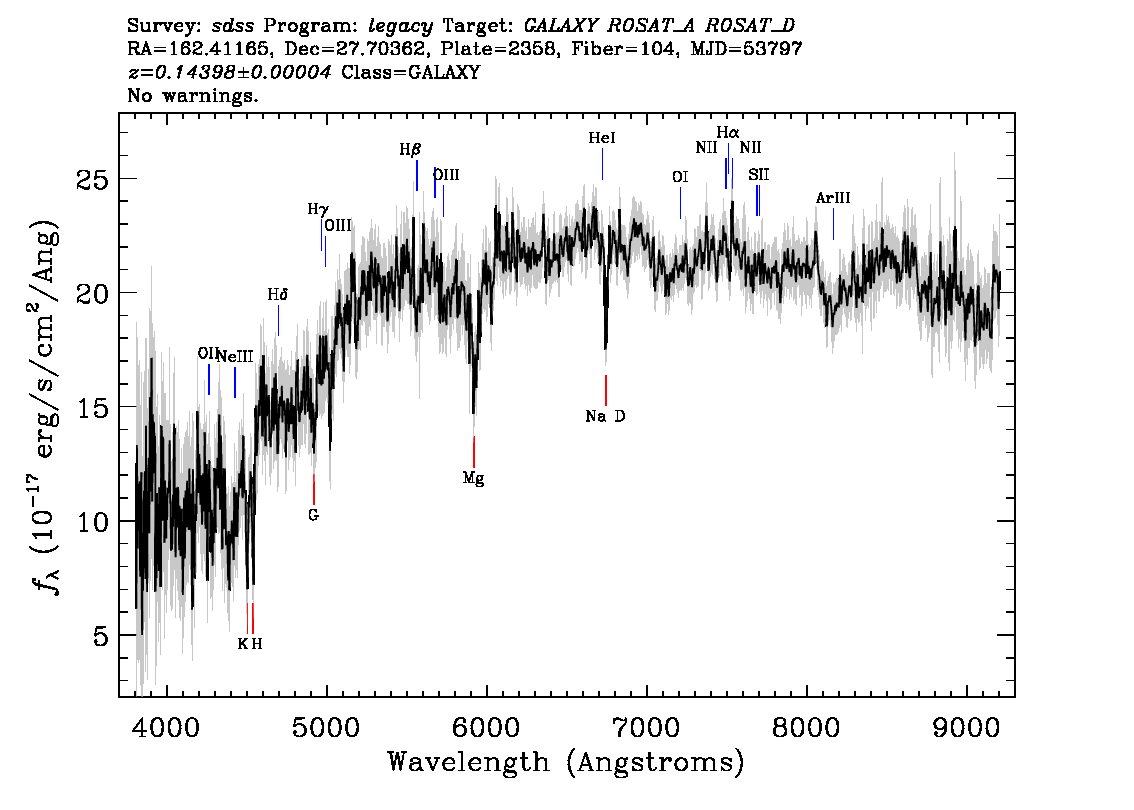}{0.5\textwidth}{(e)}
          \fig{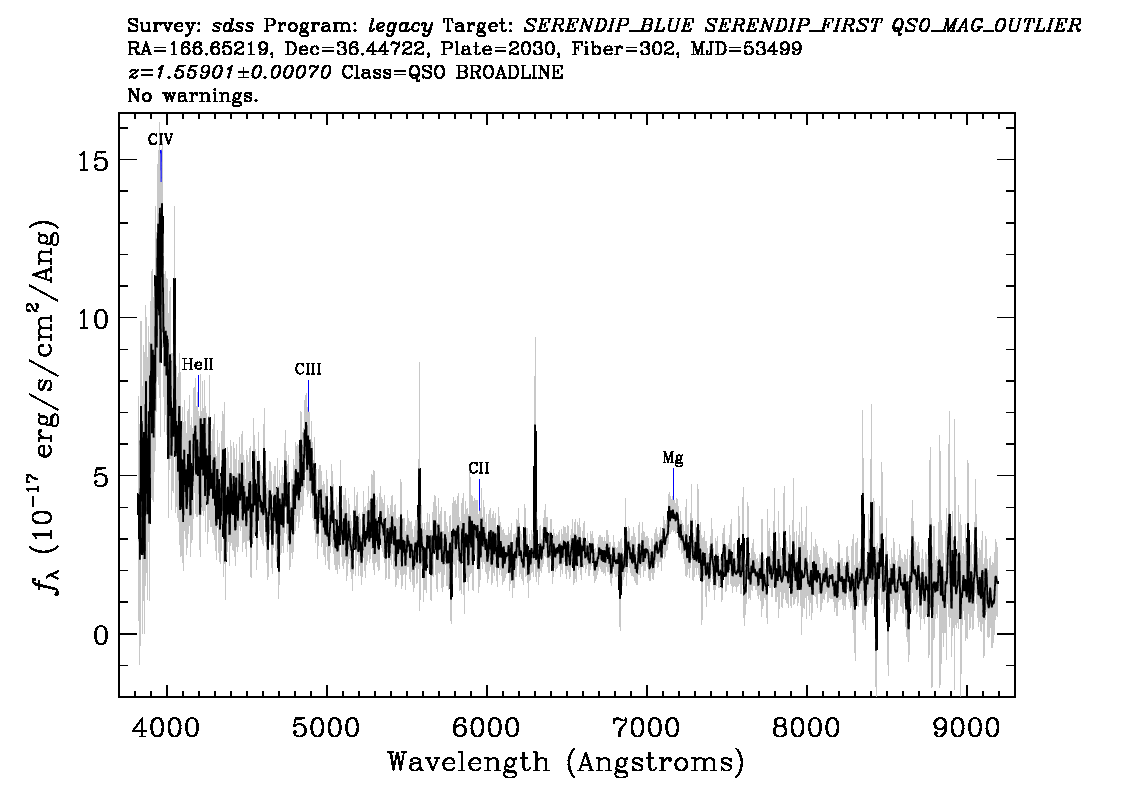}{0.5\textwidth}{(f)}
          }

\caption{The spectra for various sources as obtained from the SDSS-DR16 survey catalog are shown here. Most of our NN results are further confirmed by these spectroscopic confirmations for each subclass in the blazar category. See Table~\ref{tab:results} for details.
\label{fig:spec1}}
\end{figure*}

\begin{figure*}
          \gridline{\fig{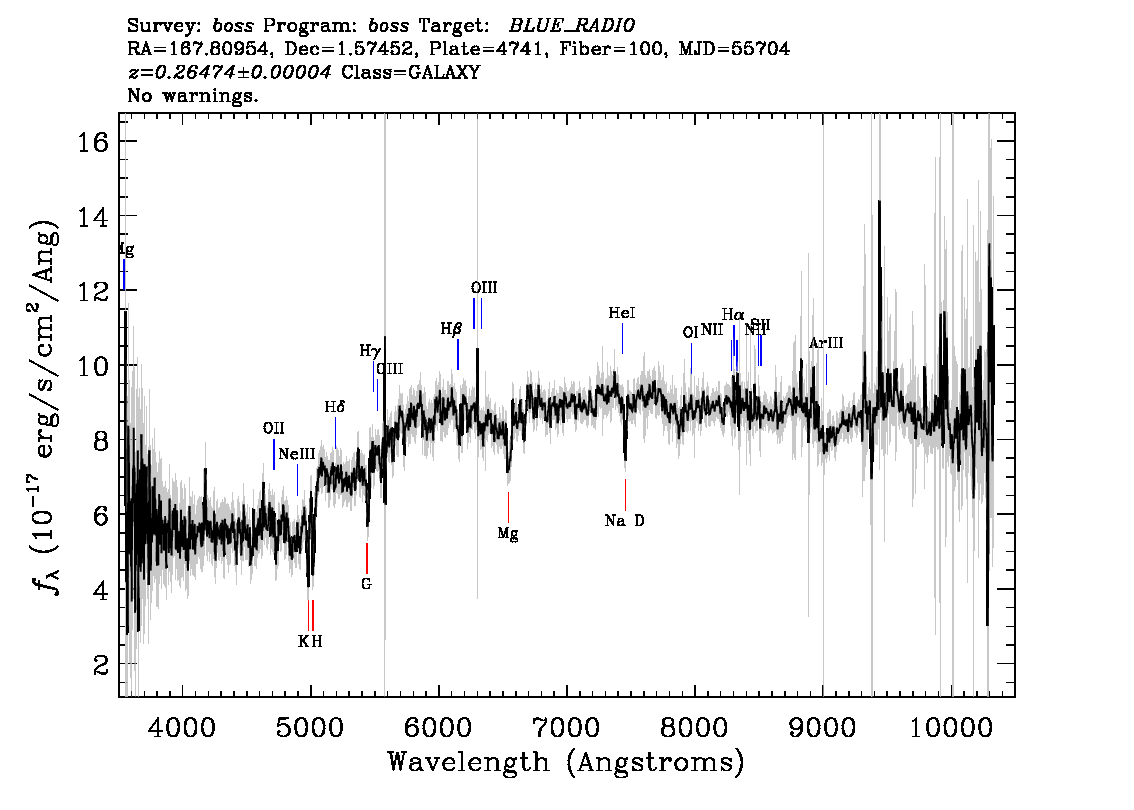}{0.5\textwidth}{(a)}
          \fig{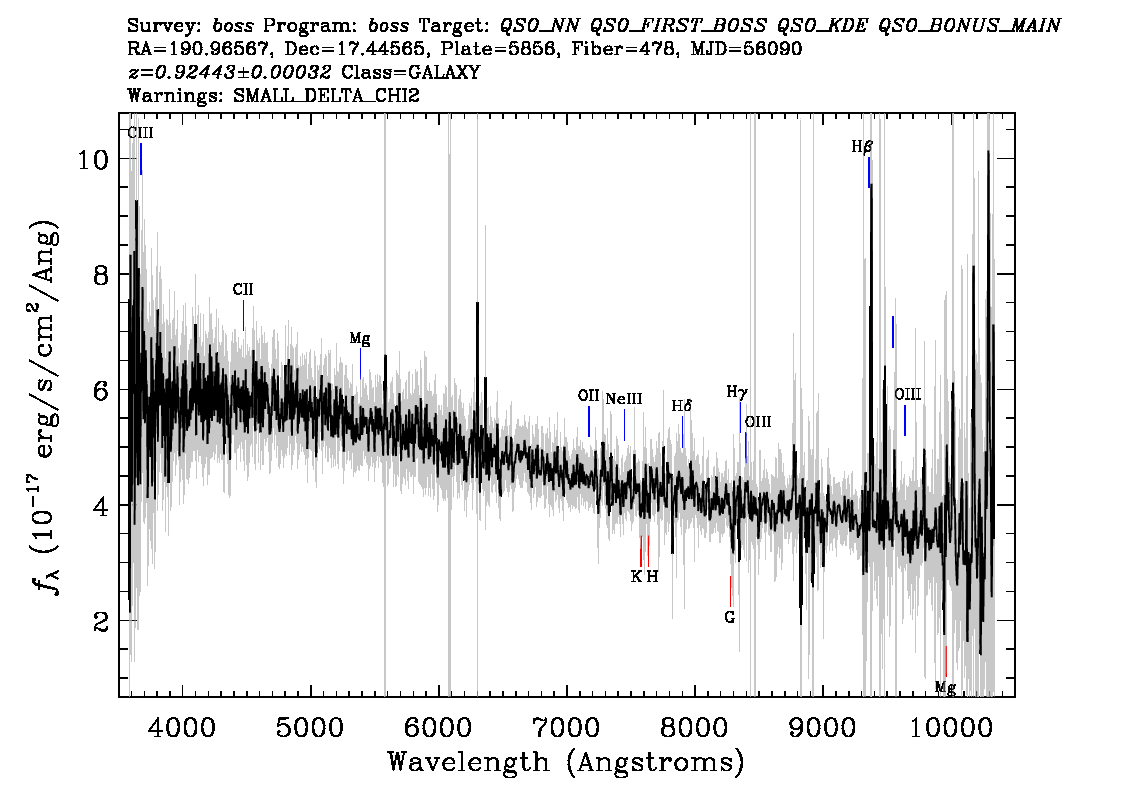}{0.5\textwidth}{(b)}
          }
          \gridline{\fig{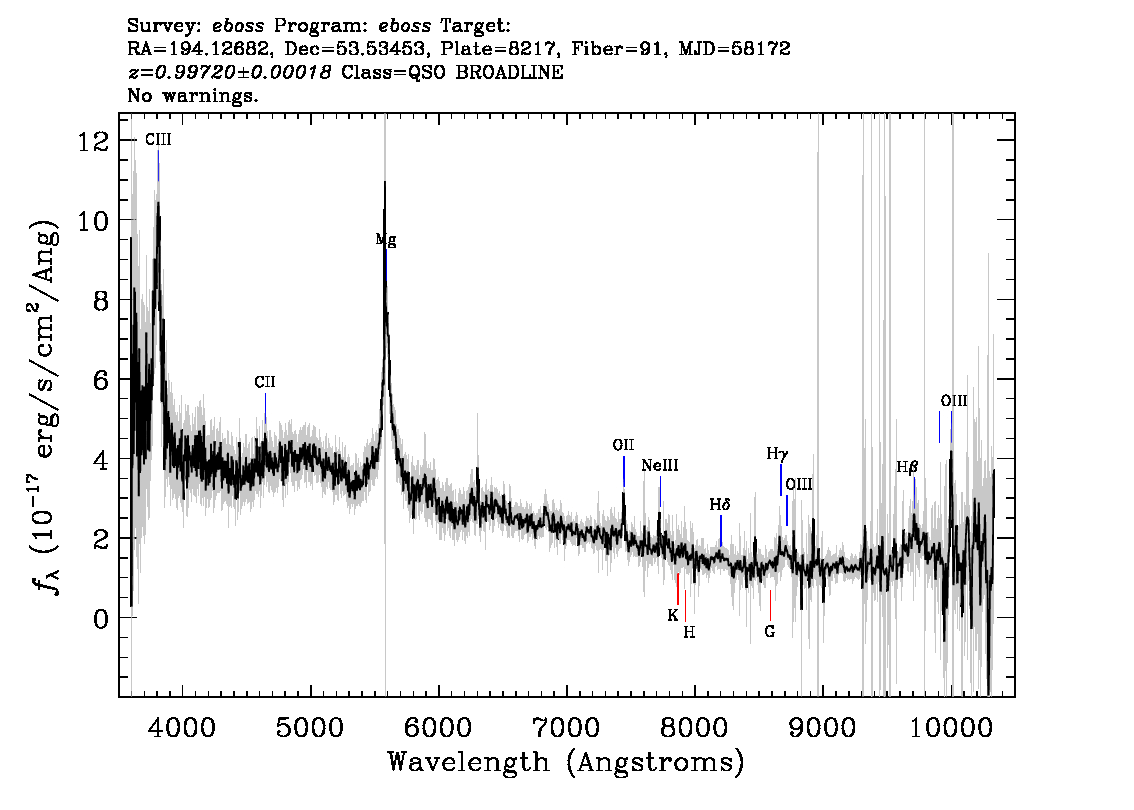}{0.5\textwidth}{(c)}
          \fig{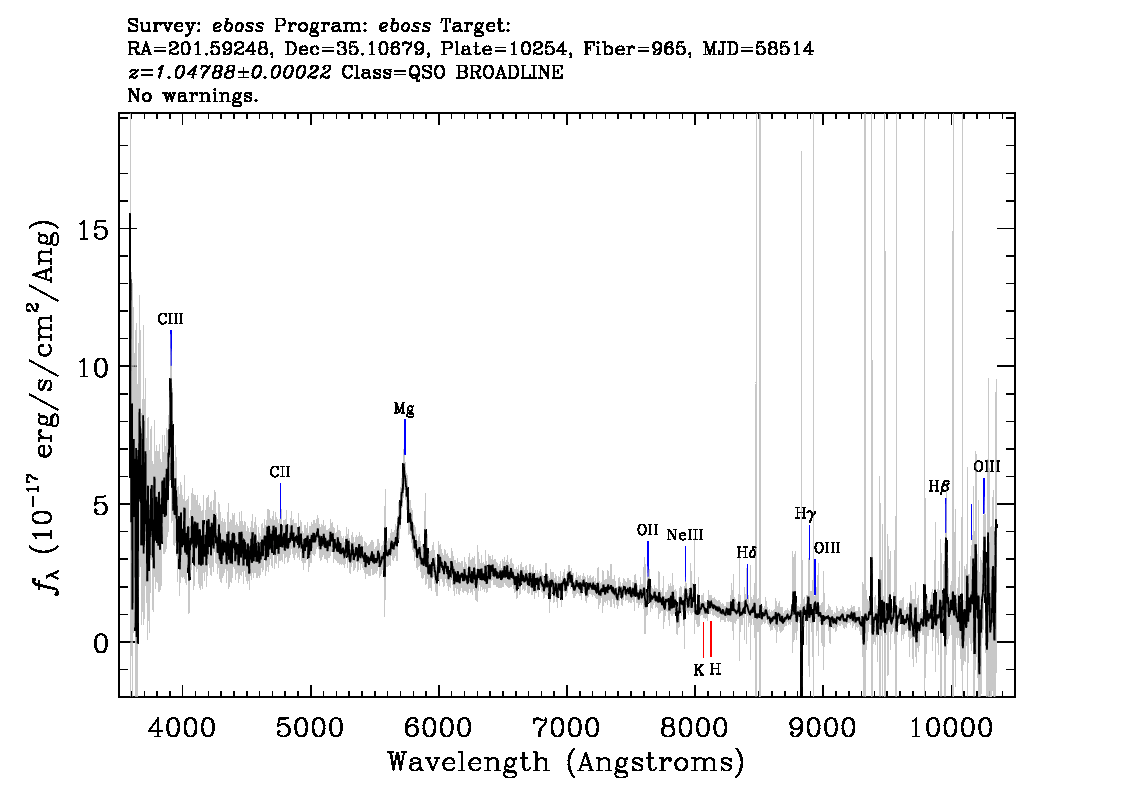}{0.5\textwidth}{(d)}
          }
          \gridline{\fig{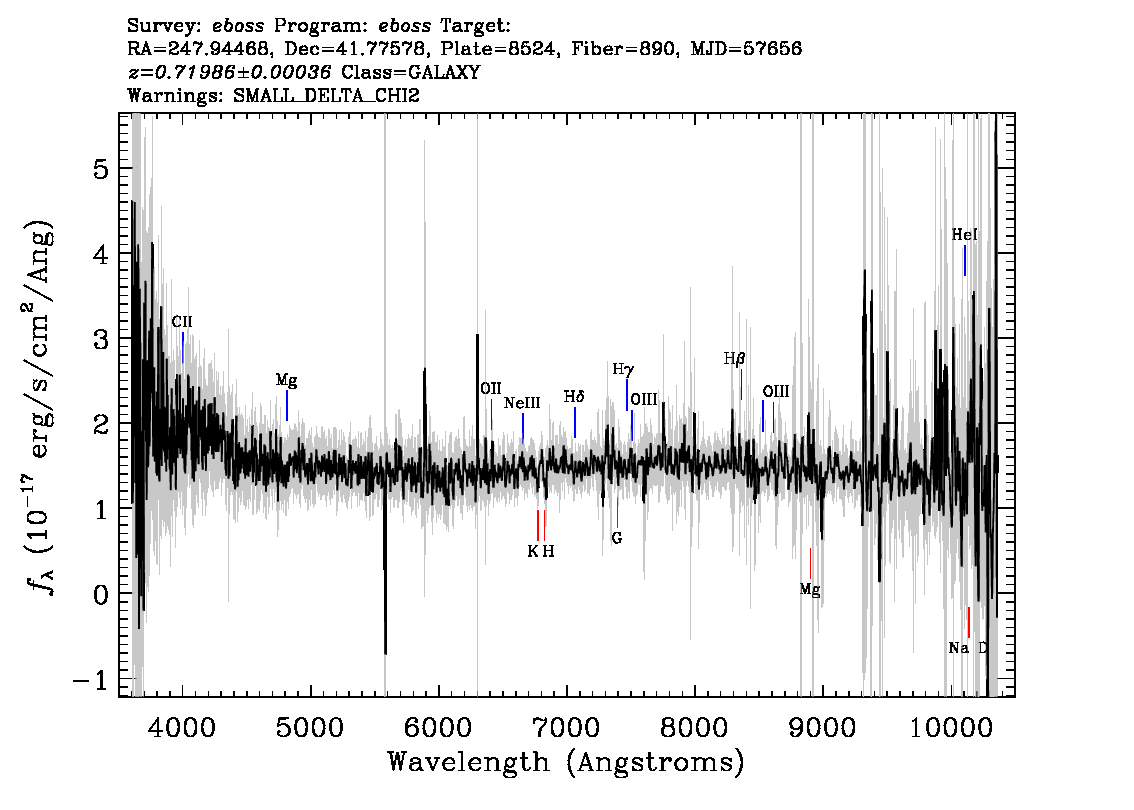}{0.5\textwidth}{(e)}
          \fig{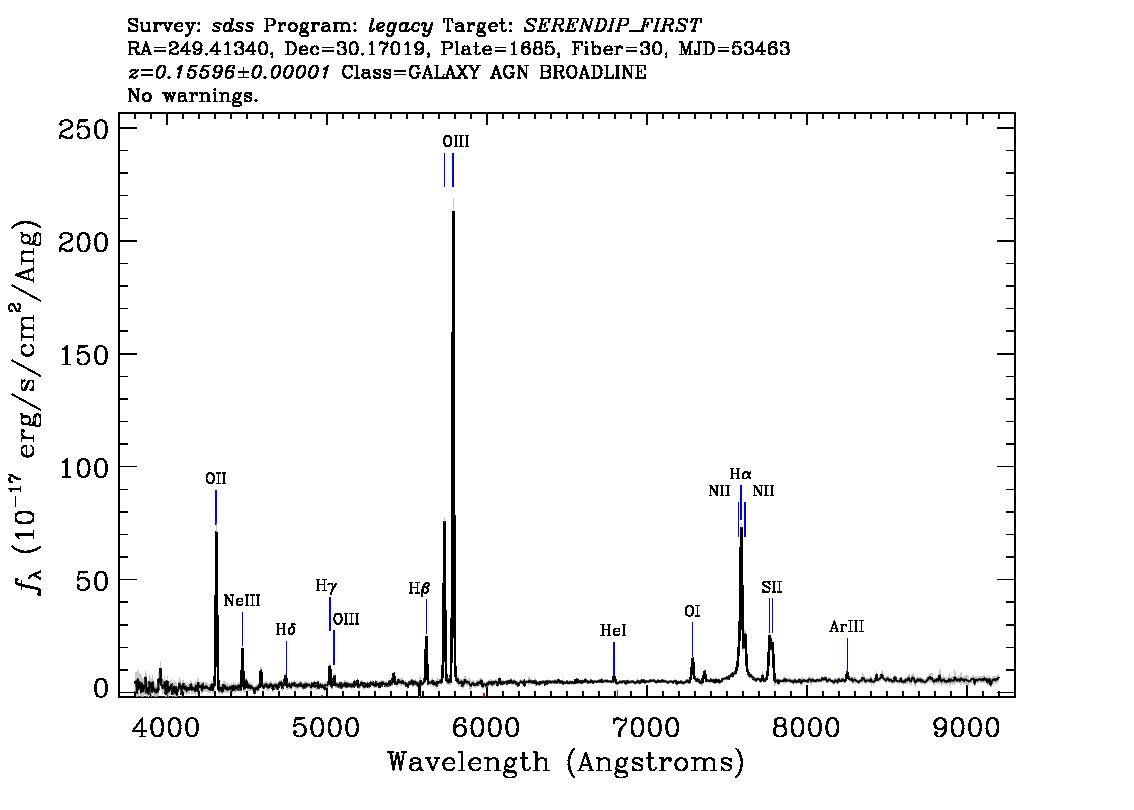}{0.5\textwidth}{(f)}
          }

\caption{The spectra for various sources as obtained from the SDSS-DR16 survey catalog are shown here. Most of our NN results are further confirmed by these spectroscopic confirmations for each subclass in the blazar category. One of the sources identified as FSRQ according to NN has been found to be a Seyfert type I (Fig~\ref{fig:spec2}f). See Table~\ref{tab:results} for more details. 
\label{fig:spec2}}
\end{figure*}

\begin{figure*}
          \gridline{\fig{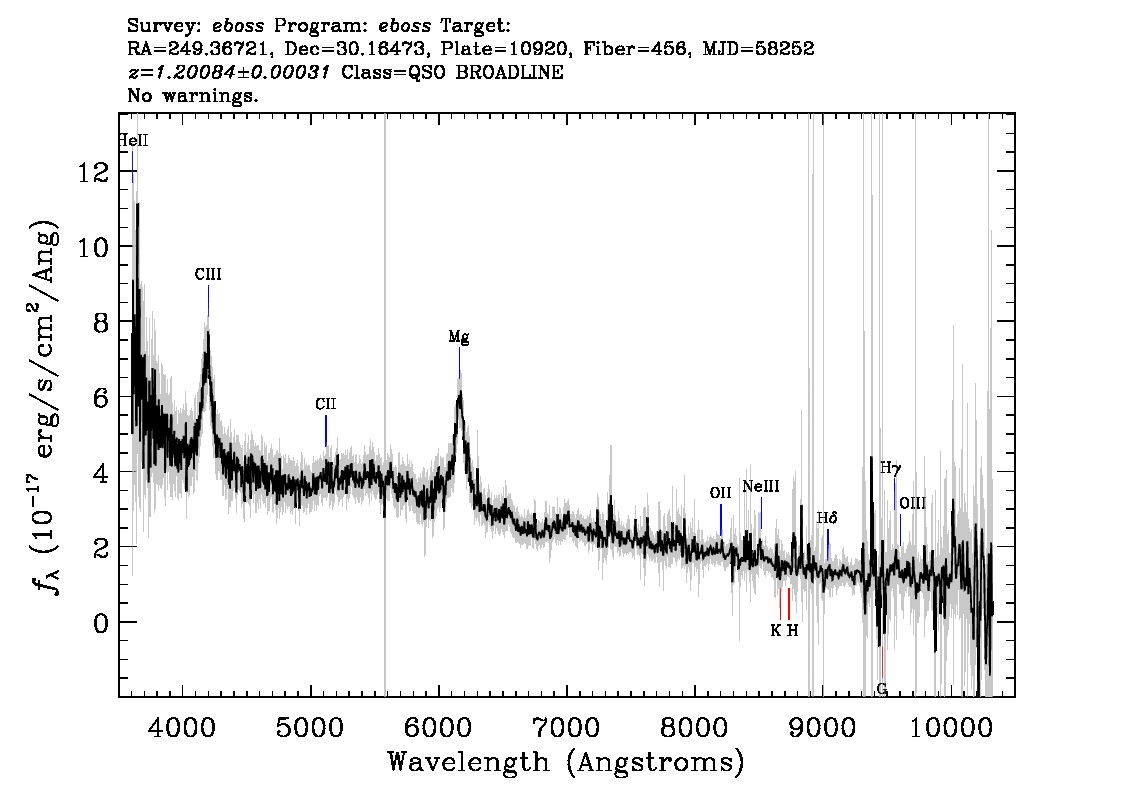}{0.5\textwidth}{(a)}
          \fig{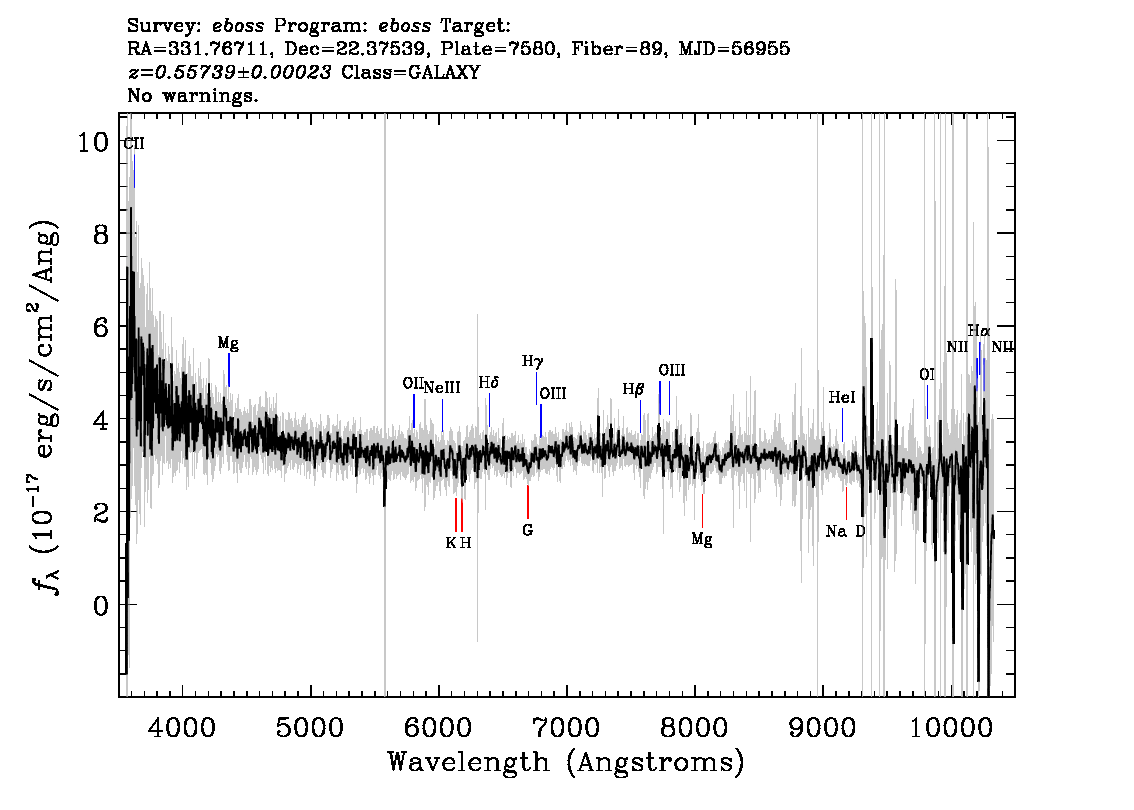}{0.5\textwidth}{(b)}
          }
          \gridline{\fig{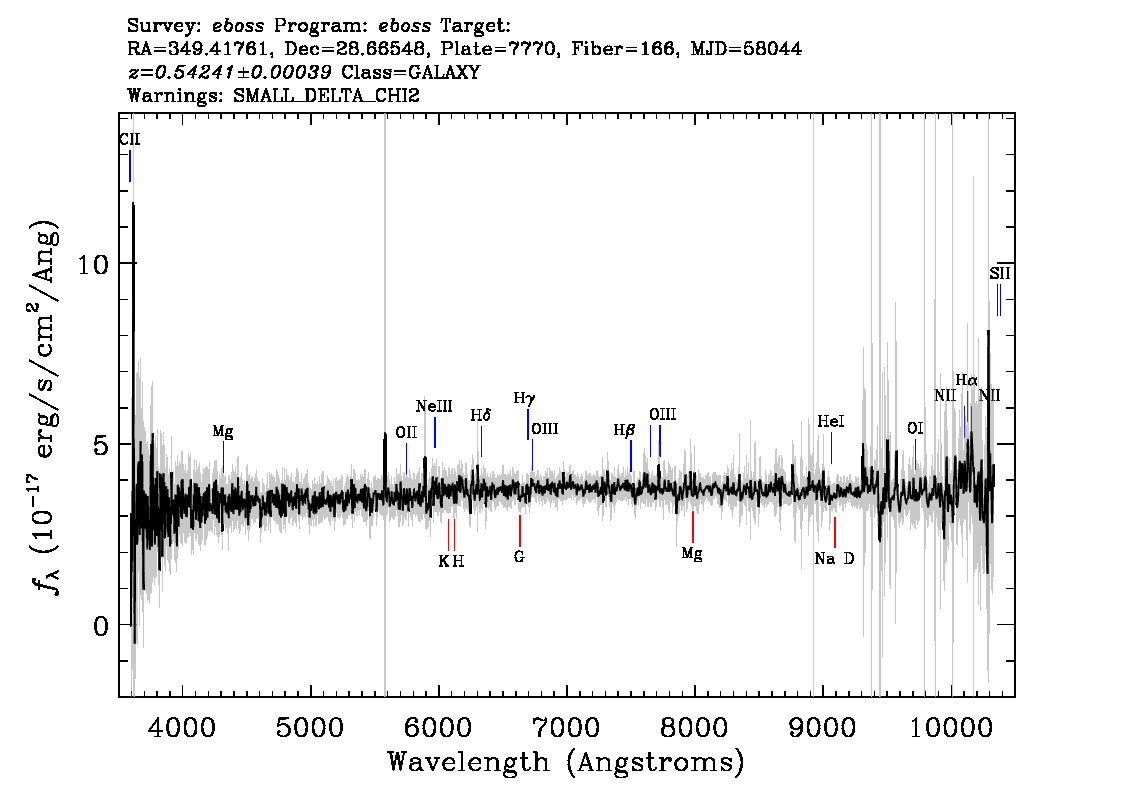}{0.5\textwidth}{(c)}
          }
          \caption{
The spectra for various sources as obtained from the SDSS-DR16 survey catalog are shown here. Most of our NN results are further confirmed by these spectroscopic confirmations for each subclass in the blazar category. See Table~\ref{tab:results} for details.
\label{fig:spec3}}
\end{figure*}

This work identifies and classifies 4 FSRQs and 50 BL Lacs from the 112 blazar candidates identified through X-ray counterpart searches among the 4FGL-DR3 unassociated sources by \citet{Kerby2021}. In addition, upon using SIMBAD and SDSS catalogs to conduct further searches for any known sources at the UV/Opt positions of these sources, we found more BL Lacs, FSRQs, radio galaxies as well as a CV and two Seyfert type I galaxies. Some of these identifications are based on the spectroscopic data taken by various studies as shown in Table~\ref{tab:results} and therefore redshift estimates are provided, while the others employed a photometric method to provide the classification as well as the redshifts. The latter work was conducted by \citet{Chang2019} which created a high-synchrotron peaked blazars catalog for various sources which are coincidentally found in our sample. It should be noted that while our NN method might not be efficient at finding each and every FSRQ or BL Lac (due to a smaller training sample size primarily), the accuracy of each identification is above 99\%, as described earlier and is seen in Figure~\ref{fig:NNCcdf}. Out of the 15 sources for which spectra were available, Figure~\ref{fig:spec1}b, ~\ref{fig:spec1}e, ~\ref{fig:spec2}a are identified as galaxies in the SDSS catalog and these appear to have signatures of lyman break galaxies. 
\begin{figure}[t]
    \centering
    \includegraphics[width=\columnwidth]{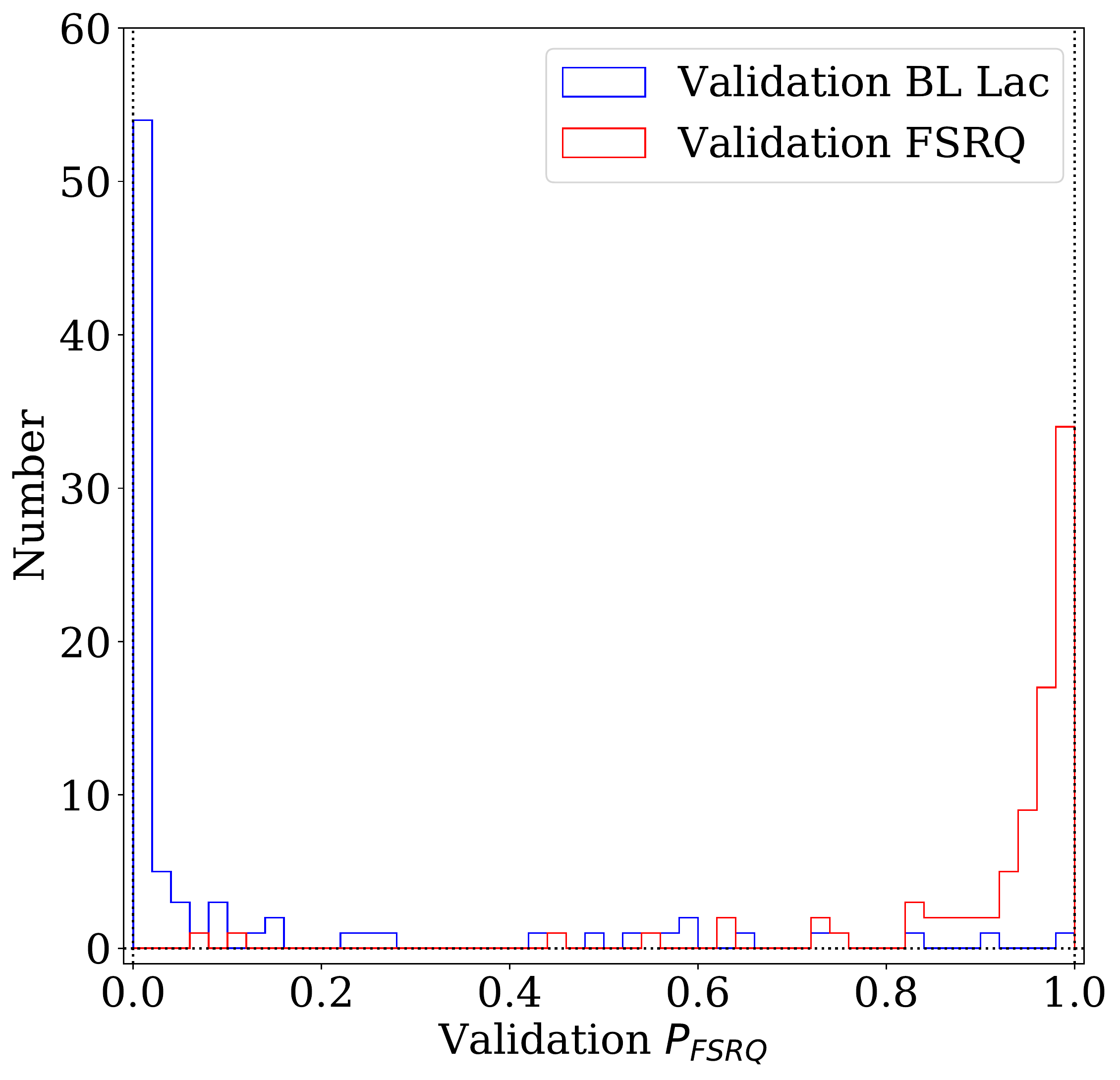}
    \caption{A histogram of the validation FSRQ (red) and BL Lac (blue) $P_{FSRQ}$ scores using the fully trained NNC.}
    \label{nnchisto}
\end{figure}
In addition, one of the two Seyfert type I sources had an optical spectrum available in the SDSS catalog. See Figure~\ref{fig:spec2}f. Figure~\ref{fig:spec1}c,~\ref{fig:spec1}f, ~\ref{fig:spec2}c,~\ref{fig:spec2}d and ~\ref{fig:spec3}a clearly represent the FSRQ population of the blazar class and the rest are BL Lacs. Although our source sample would ideally be comprised of only blazar candidates, a few non-blazar sources were found to be in X-ray and UV positional coincidence. 4FGL J1513.0-3118 is found to be a positional match with a rotationally variable star (possibly a new gamma-ray binary); 4FGL J1910.8+2856 with a cataclysmic variable star; and 4FGL J2351.4-2818 with a bright Galaxy in a cluster. Further investigations of these sources at both high and low energy regimes are required before any firm conclusions, which are out of the scope for this work. Overall, this work yields 48\% classification of blazar candidates into BL Lacs and FSRQs such that the former represent 45\% and the latter 3\% of the total classification. The rest 52\% are blazars of ambiguous type based on our work, as shown in Table~\ref{tab:results}. This work is a substantial step taken towards the completeness of the gamma-ray emitting blazar class in the Fermi catalog. Furthermore, this work along with \citep{Kerby2021} has identified/classified various sources among the unassociated Fermi catalog using the X-ray, UV/optical, and IR, along with the gamma-ray data.

\begin{acknowledgments}
The work presented in this manuscript was made possible due to the NASA research grants \textcolor{blue}{80NSSC20K0911} and \textcolor{blue}{80NSSC20K1577}. This publication made use of data products from the Wide- field Infrared Survey Explorer, which is a joint project of the University of California, Los Angeles, and the Jet Propulsion Laboratory/California Institute of Technology, funded by the National Aeronautics and Space Administration. The data employed in this research were extracted from the High Energy Astrophysics Science Archive Research Center (HEASARC), which is a service of the Astrophysics Science Division at NASA/GSFC. 
\end{acknowledgments}
\facilities{Fermi, Swift(XRT and UVOT), WISE}

\software{Astropy \citep{astropy:2013}, numpy \citep{numpy}, scikitlearn \citep{scikit-learn}, Topcat \citep{taylor2005}}

\clearpage
\bibliography{bib_bll_fsrq}{}
\bibliographystyle{aasjournal}
\input{tables/table_training}
\input{tables/table_unassoc}

\input{tables/table_results}

\end{document}

%% file: tables/table_training.tex
\startlongtable
\begin{deluxetable*}{cclcccccc}
\tablecaption{The 1098 sources used for training data containing BL Lacs and Flat Spectrum Radio Quasars  \label{tab:bzrs}}
\tablewidth{0pt}
\tablehead{
\colhead{Fermi} & \colhead{log F$_\gamma$} & \colhead{$\Gamma_\gamma$} & \colhead{Variability} & \colhead{Curvature} & \colhead{log F$_X$/F$\gamma$} & \colhead{log F$_V$/F$\gamma$} & \colhead{w1-w2} & \colhead{w2-w3} \\
\hline
\colhead{4FGL} & \colhead{}  & \colhead{} & \colhead{log} & \colhead{log} & \colhead{} & \colhead{} & \colhead{} & \colhead{} }
\startdata
  \hline
    &  &  & BL Lacs & & & & & \\ 
  \hline
J0001.2$-$0747 & -11.04 & 2.1 & 0.27 & 1.52 & -1.08 & -4.96 & 0.95 & 2.59\\
J0006.3$-$0620 & -11.81 & 2.13 & -0.06 & 1.31 & 0.06 & -4.18 & 1.14 & 2.99\\
J0008.0+4711 & -10.74 & 2.05 & 0.66 & 0.81 & -1.2 & -5.4 & 0.94 & 2.42\\
J0008.4$-$2339 & -11.63 & 1.66 & 0.47 & 1.14 & 0.35 & -4.37 & 0.55 & 1.93\\
J0009.1+0628 & -11.27 & 2.1 & 0.08 & 1.12 & -0.64 & -4.82 & 0.94 & 2.76\\
J0009.3+5030 & -10.61 & 1.99 & 0.67 & 2.17 & -1.67 & -4.72 & 0.63 & 2.21\\
J0013.9$-$1854 & -11.48 & 2.0 & 0.07 & 0.64 & 0.63 & -4.04 & 0.23 & 1.32\\
\enddata
\tablecomments{The table parameters are as follows: The logarithm of gamma-ray flux (log ${F_\gamma}$ (erg/s/cm$^2$)), Gamma-ray photon index ($\Gamma_\gamma$), the significance of the curvature in the gamma-ray spectrum (Curvature), the year-over-year gamma-ray variability index (Variability) are derived from the 4FGL-DR3 catalog. The last four columns correspond to the logarithm of X-ray to gamma-ray flux ratio ($\log{F_X/F_\gamma}$), the logarithm of V-band to gamma-ray flux ratio ($\log{F_V/F_\gamma}$) , WISE color index 1(w1-w2) and WISE color index 2 (w2-w3).}
\end{deluxetable*}

%% file: tables/table_unassoc.tex
\begin{longrotatetable}
\begin{deluxetable*}{cclclcccccc}
\tablecaption{The Source Sample: the 112 blazar candidates obtained from \citet{Kerby2021}\label{tab:sample}}
\tablewidth{0pt}
\tablehead{
\colhead{Fermi} & \colhead{Swift-XRT}  & \colhead{Swift-UVOT} & \colhead{log F$_\gamma$} & \colhead{$\Gamma_\gamma$} & \colhead{Variability} & \colhead{Curvature} & \colhead{F$_X$/F$_\gamma$} & \colhead{F$_V$/F$\gamma$} & \colhead{w1-w2} & \colhead{w2-w3} \\
\colhead{4FGL} & \colhead{SwXF4}  & \colhead{SwUVF4} & \colhead{} & \colhead{} & \colhead{log} & \colhead{log} & \colhead{log} & \colhead{log} & \colhead{} & \colhead{}
}
\startdata
J0004.4$-$4001 & J000434.1-400036 & J000434.2-400035 & -11.57 & 2.15 & -0.14 & 1.54 & -0.84 & -4.04 & 0.48 & 2.38\\
J0025.4$-$4838 & J002536.8-483808 & J002536.9-483810 & -11.69 & 2.14 & 0.26 & 1.16 & -0.62 & -4.5 & 0.6 & 1.69\\
J0026.1$-$0732 & J002611.9-073115 & J002611.6-073115 & -11.77 & 1.94 & 0.03 & 0.95 & 0.42 & -4.44 & 0.55 & 2.22\\
J0031.5$-$5648 & J003135.1-564640 & J003135.1-564640 & -11.81 & 1.99 & 0.13 & 1.16 & -0.59 & -4.68 & 0.32 & 2.52\\
J0037.2$-$2653 & J003729.6-265043 & J003729.5-265045 & -12.0 & 2.34 & 0.21 & 0.81 & -0.33 & -4.44 & 0.99 & 3.19\\
J0057.9+6326 & J005758.1+632642 & J005758.1+632642 & -11.5 & 1.71 & 0.3 & 1.16 & -0.05 & -4.58 & 0.31 & 2.63\\
J0058.3$-$4603 & J005806.4-460417 & J005806.3-460419 & -12.03 & 2.05 & -0.22 & 1.21 & 0.51 & -4.72 & 0.11 & 2.52\\
J0118.3$-$6008 & J011824.0-600753 & J011824.0-600753 & -11.73 & 2.18 & 0.3 & 1.49 & -1.01 & -4.45 & 0.68 & 1.94\\
\enddata
\tablecomments{All the parameters are described in Table 1.}
\end{deluxetable*}
\end{longrotatetable}

%% file: tables/table_results.tex
\startlongtable
\begin{deluxetable*}{ccccc}
\tablecaption{The association results based on the neural network probabilities for all the 4FGL unassociated sources \label{tab:results} }
\tablewidth{\columnwidth}
\tablehead{
\colhead{4FGL Name} & \colhead{UVOT} & \colhead{P\_{fsrq}} & \colhead{SIMBAD/SDSS} & \colhead{redshift} 
}
\startdata
\hline
 & & &\textbf{FSRQ candidates from NN results} &\\
 \hline
J1008.2$-$1000&J100802.5$-$095918&0.993&QSO&1.66\citep{Krogager2018}\\
J1008.2$-$1000&J100749.4$-$094912&0.993&\nodata&\nodata \\
J1637.5+3005&J163728.1+300953&0.992&QSO&1.2; Fig~\ref{fig:spec3}a\\
J1637.5+3005&J163739.2+301013&0.991&Seyfert I&0.155; Fig~\ref{fig:spec2}f\\
\hline
 & & &\textbf{BL Lac candidates from NN results} &\\
 \hline
J0026.1$-$0732&J002611.6$-$073115&0.007&blazar candidate&0.5\tablenotemark{ph} \\
J0031.5$-$5648&J003135.1$-$564640&0.008&\nodata&\nodata \\
\enddata
\tablenotetext{*}{ The redshift values provided by the SDSS$-$DR16 and SIMBAD catalogs are different; 0.94 and 2.14 respectively. In addition, the redshift estimate has varied from 0.5 to 2.14 in different data releases for SDSS. The value provided in this table corresponds to the latest data release, DR16.}
\tablenotetext{ph}{\,\,photometric redshift from the High$-$Synchrotron Peaked blazars catalog by \citet{Chang2019}}
\tablenotetext{bin}{\,\,this source is classified as binary source in the most recent third data release 4FGL catalog (4FGL$-$DR3) \citep{2022-4FGL}}
\end{deluxetable*}